\documentclass[final,5p,times,twocolumn]{elsarticle}

\usepackage{amsmath,amssymb,amsthm,bm,widetext}
\usepackage{tikz}
\usepackage{graphicx}
\usepackage{standalone}
\usepackage[utf8]{inputenc}

\newcommand{\angstrom}{\text{\normalfont\AA}}

\journal{Journal of Computational Physics}

\begin{document}

\begin{frontmatter}



\title{Massively parallel symplectic algorithm for coupled magnetic spin dynamics and molecular dynamics}


\author[snl1]{J.~Tranchida}\corref{cor1}
\ead{jtranch@sandia.gov}
\author[snl1]{S.J.~Plimpton}
\ead{sjplimp@sandia.gov}
\author[cealr1]{P.~Thibaudeau}
\ead{pascal.thibaudeau@cea.fr}
\author[snl1]{A.P.~Thompson}
\ead{athomps@sandia.gov}

\cortext[cor1]{Corresponding author}

\address[snl1]{Multiscale Science Department, Sandia National Laboratories, P.O. Box 5800, MS 1322, Albuquerque, NM 87185}
\address[cealr1]{CEA-DAM Le Ripault, BP 16, F-37260 Monts, France }

\begin{abstract}

A parallel implementation of coupled spin--lattice dynamics in the
LAMMPS molecular dynamics package is presented.  The equations of
motion for both spin only and coupled spin--lattice dynamics are first
reviewed, including a detailed account of how magneto-mechanical
potentials can be used to perform a proper coupling between spin and
lattice degrees of freedom.  A symplectic numerical integration
algorithm is then presented which combines the Suzuki--Trotter
decomposition for non-commuting variables and conserves the geometric
properties of the equations of motion.  The numerical accuracy of the
serial implementation was assessed by verifying that it conserves the
total energy and the norm of the total magnetization up to second
order in the timestep size.  Finally, a very general parallel
algorithm is proposed that allows large spin--lattice systems to be
efficiently simulated on large numbers of processors without degrading
its mathematical accuracy.  Its correctness as well as scaling
efficiency were tested for realistic coupled spin--lattice systems,
confirming that the new parallel algorithm is both accurate and
efficient.

\end{abstract}


\end{frontmatter}


\section{Introduction}\label{sec:intro}

Magnetization processes encapsulate several fundamental attributes of
ferromagnetic materials including inherent hysteresis and constitutive
nonlinearities due to the cooperative domain structure of these
materials.  However, magneto-mechanical properties of the materials
provide actuator and sensor capabilities that enable design of
contemporary and future smart devices\cite{smith_smart_2005}.  In
general these effects are highly complex. As a result, the design of
devices such as transducers may require manipulation of multiple
distinct mechanisms acting in concert.  Magneto-mechanical effects or
magneto-elasticity refers to a family of mechanisms in which (i) an
applied stress causes magnetic moments to rotate, thus changing the
magnetization and (ii) an applied field causes the rotation of
magnetic moments, thus generating strain in the material.  While these
effects are often viewed as limitations, many technological
applications rely on them to achieve novel and unique functionalities.
In both cases, their understanding is of fundamental
importance. Accurate numerical simulations of magneto-elasticity
fulfill two essential roles (i) enabling the underlying assumptions of
competing theories to be objectively tested and (ii) predicting and
interpreting experimental observations.

It is well known that the response of a magnetic material subjected to
an external field has its root in the collective interaction of the
electronic magnetic moments of the atoms.  This is particularly true
when the external field induces a mechanical stress, because of the
associated slow relaxation processes \cite{rossi_dynamics_2005}.  
Interactions between spin and
lattice systems can also occur when the external stimulus is applied
to the electronic system.  A salient example can be seen in the
Beaurepaire's experiments \cite{beaurepaire_ultrafast_1996}.  By
probing the magnetic and mechanical relaxations after applying an
electronic stimulus, the intimate couplings between electrons, spins,
and lattice (nuclear coordinates) were characterized.

\begin{figure}[h]
\centering
\begin{tikzpicture}
\draw[line width=2pt] (0.0,0.0) circle (1.0);
\node[align=left] (e) at (0.0, 0.0) {Electron \\ $\tau_{\rm e} \approx 1$~fs };
\draw[line width=2pt] (4.0,0) circle (1.0);
\node[align=left] (l) at (4.0, 0.0) {~Lattice \\ $\tau_{\rm l} \approx 1$~ps};
\draw[line width=2pt] (2.0,2.83) circle (1.0);
\node[align=left] (s) at (2.0, 2.83) {~~~Spin \\ $\tau_{\rm s} \approx 1$~ps};
\draw[line width=2pt, <->] (1.0,0.0) -- (3.0,0.0);
\draw[line width=2pt, <->] (0.5,0.87) -- (1.29,2.12);
\draw[line width=2pt, <->] (3.5,0.87) -- (2.707,2.12);
\node (el) at (2.0, -0.5) {$\tau_{\rm e-l} \approx 1$~ps};
\node (es) at (-0.2, 1.5) {$\tau_{\rm e-s} \approx 0.1$~ps};
\node (sl) at (4.2, 1.5) {$\tau_{\rm s-l} \approx 100$~ps};
\end{tikzpicture}
\caption{Schematic overview of the three subsystems with their
  associated characteristic relaxation times and coupling
  parameters.} \label{fig:3T}
\end{figure}
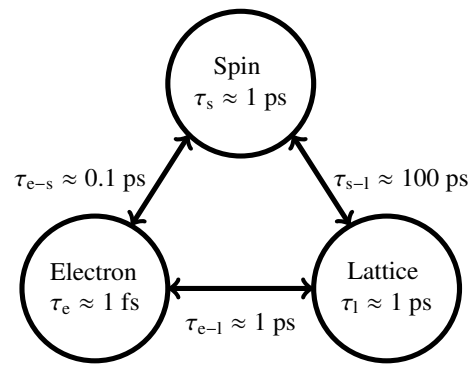

Fig.~\ref{fig:3T} shows characteristic timescales of spin, lattice,
and electron dynamics that have been experimentally measured and
reported.

Because the electronic relaxation time is much smaller than all the
other timescales, the electron subsystem can be considered to relax
instantaneously to an equilibrium state in response to changes in the
atom positions or the magnetic spin moments.  This allows us to focus
on the coupled dynamics of spin and lattice, and their associated
relaxation processes.

To study spin--lattice relaxation processes in full detail, an
accurate technique is the Time--Dependent--Spin--Density--functional
theory (TD-SDFT) \cite{qian_dynamical_2002}.  Within TD-SDFT, local
effects including spin-relaxation and mechanical constants can be
studied with high accuracy, and magneto-mechanical potentials can be
evaluated \cite{restrepo_full_2012,ozdemir_kart_elastic_2010}.

When collective, large--scale behaviors of atomic motions are of
interest, molecular dynamics simulation (MD) is an extremely powerful
technique to solve the classical many-body problem. This is
particularly true when the long-time dynamics of large systems are
studied, which is often necessary to achieve physical
equilibrium. Consequently, for systems that obey the ergodic
hypothesis, the evolution of MD trajectories in phase space is useful
to determine macroscopic thermodynamic properties of the Hamiltonian
system. The time averages of ergodic systems generate accurate
micro-canonical ensemble averages \cite{mcquarrie_statistical_1976}.

However, the properties of magnetic materials emerge from the combined
effects of physical interactions operating on multiple lengthscales
(i) extremely localized effects, such as the exchange interaction
responsible for local alignment of electronic magnetic moments (spins)
on neighboring atoms, (ii) long-range dipolar interactions between
large assemblies of spins, and (iii) long-range ordering of spins
leading to magnetic domains and domain--walls
\cite{aharoni_introduction_1996}.  For these reasons, methods which
enable both the definition of local interactions and the possibility
of simulating large systems are highly desirable.  In this regard,
quantum local methods, like TD-SDFT (very accurate description of the
dynamics of hundreds atoms for picoseconds), and semi-classical
methods, like molecular or spin dynamics (less accurate, but enabling
the simulation of millions of atoms for nanoseconds) fill
complementary roles.

To achieve a connection between the two methodologies, Antropov
\emph{et al.} first proposed a transition from a general
quantum-mechanical formulation to classical equations of motion (EOM),
first for spin--only systems \cite{antropov_ab_1995} and then for
coupled spin--lattice dynamics \cite{antropov_spin_1996}. This
methodology starts from general principles and introduces an adiabatic
approach that considers the orientation of the local magnetic moments
to be slowly varying relative to their magnitudes. Because the
orientation of the magnetization density is introduced as a collective
variable in density functional theory, the derived EOM for spin--only
systems can be combined with those of first-principles MD to derive a
consistent treatment of spin-lattice interactions.

For spin--only dynamics, lattices of fixed atoms are constructed, and
precession EOM for each classical spin vector are integrated.
\cite{skubic_method_2008,eriksson_atomistic_2017}.  This methodology,
which is special case of spin--lattice simulations, proved to be an
efficient way to study many magnetic phenomena, such as ultrafast
magnetization reversal \cite{radu_transient_2011}, or the
configuration of topological spin structures like skyrmions
\cite{dupe_tailoring_2014}.  In addition, this approach can be used to
obtain statistical averages over a large number of spins
\cite{tranchida_functional_2016} to generate effective magnetization
dynamics.

Without considering the full machinery of first-principles methods and
motivated by the fact that the characteristic time scales for the
motion of spin and lattice degrees of freedom are very slow compared
to those of the electron subsystem (see Fig.~\ref{fig:3T}), the
dynamics of both spin and lattice variables can be treated using
effective, parametrized potentials. We refer to this approach as
SD--MD.  In a sense, this allows a separation of the spin and lattice
variables from their electronic counterparts, and derivation of
coupled EOM for the spin--lattice system.

A few implementations of this SD--MD methodology have been published.
Ma \emph{et al.} proposed the application of SD--MD to the study of
magnetostriction properties of iron
\cite{ma_large-scale_2008,ma_spilady:_2016}.  They were primarily
concerned with demagnetization experiments, phase transitions, and the
impact of temperature on the spin-lattice system.  Because of this,
they considered only the exchange interaction and neglected
anisotropic energy originating from the spin-orbit coupling.

Beaujouan \emph{et al.} added both one-site and two-site anisotropic
energies and applied SD--MD to the study of cobalt nano-systems,
recovering magnetostriction properties
\cite{beaujouan_anisotropic_2012}.  More recently, Perera \emph{et
  al.} \cite{perera_reinventing_2016,perera_collective_2017} explored
this coupling method to investigate the mutual influence of phonons
and magnons on their respective frequency spectra and lifetimes in
ferromagnetic bcc iron.

In this paper we describe a general implementation of the SD--MD
methodology in the LAMMPS molecular dynamics package
\cite{plimpton_lammps-large-scale_2007}. It integrates many of the
previous published improvements, algorithms, and potentials within the
spatial decomposition framework of LAMMPS to enable large-scale
parallel simulations.  Section~\ref{sec:dynamics} presents the EOM for
spin--only and spin--lattice dynamics with magnetic and
magneto-mechanics potentials.  Details of these potentials and their
parametrization are given in \ref{app:exchange}, \ref{app:hamiltonian}
and \ref{app:mechpotential}.  Non-adiabatic spin-lattice systems
coupled to a thermal reservoir are handled using the Langevin
approach, which is described in Section~\ref{sec:temperature}.  In
order to preserve the phase space volume adiabatically during the
dynamics, numerical solution of the spin--lattice EOM requires
particular care. Symplectic geometric methods based on the
Suzuki--Trotter decomposition of the discretized equations
\cite{blanes_magnus_2009} were used to derive a particular algorithm
that is presented in Section~\ref{sec:algorithms}, along with results
demonstrating its accuracy.  An efficient parallel implementation of
this spin--lattice algorithm was developed following a synchronous
sub-lattice sectoring methodology \cite{shim_semirigorous_2005}.
Section~\ref{sec:parallel} first details this algorithm, and then
analyzes its accuracy and parallel scalability.


\section{Spin--lattice dynamics in the microcanonical ensemble}\label{sec:dynamics}

A microcanonical ensemble (NVE) defines the statistical ensemble that
is used to represent all the possible states of a closed mechanical
system with fixed total energy, composition, volume, and shape. The
equations of motion including both whole magnetics and mechanics
degrees of freedom in ths NVE ensemble are described below.

\subsection{Equations for atomistic spin dynamics}\label{subsec1:dynamics}

Following the approach of Antropov \emph{et al.}
\cite{antropov_ab_1995,antropov_spin_1996}, a lattice of atomic sites
is considered. A classical spin vector $\bm{S}_{i}$ is associated
with each site labelled by $i\in 1\dots N$, with $\bm{S}_{i} =
\hbar g_{i}\bm{s}_{i}$ where $g_{i}$ is the Land\'e factor
of the spin $i$, eventually calculated by first-principles methods,
and $\bm{s}_{i}$ is a unitary, non-dimensional vector.
Consequently the adiabatic motion in time of such atomic spin vector
$\bm{s}_{i}$ can be described by a classical model of Thomas's
precession
\cite{bacry_thomass_1962,thomas_motion_1926,thomas_i._1927}.  By
considering the fluctuation-dissipation theorem, finite temperature
effects produce friction-like damping on the classical precession
motion. This comes from the connection of a closed quantum system to
external sources \cite{weiss_quantum_2012,wieser_description_2015}.
This can be taken into account phenomenologically by the
Landau--Lifshitz--Gilbert equation
\cite{gilbert_phenomenological_2004}

\begin{equation}
 \frac{d \bm{s}_{i}}{dt} = \frac{1}{\left(1+\lambda^2 \right)}
 \left(\bm{\omega}_{i} - \lambda\, \bm{\omega}_{i}
 \times\bm{s}_{i}\right)\times \bm{s}_{i} , \label{LLG}
\end{equation}
with $\lambda$ a purely transverse damping constant, and
$\bm{\omega}_{i}$ the analog of a spin force applied to the spin
$i$, and defined as
\begin{equation}
  \bm{\omega}_{i} = -\frac{1}{\hbar} \frac{\partial
    \mathcal{H}_{mag}}{\partial \bm{s}_{i}},
\end{equation}

with $\hbar=6.582 \times 10^{-7}$ eV.(rad.THz)$^{-1}$ the reduced
Planck constant, and $\mathcal{H}_{mag}$ the spin Hamiltonian of
the magnetic system. In the formalism of the Nos\'{e}--Hoover
thermostat, $\lambda$ is not a constant and may vary in time
\cite{antropov_ab_1995}, or can even be determined by first-principles
methods \cite{ebert_ab_2011}.  Extension to chains of thermostats can
also be considered as a faster numerical way to reach thermal
equilibrium \cite{thibaudeau_thermostatting_2012}.

A simple expression for the Hamiltonian of $N$ interacting spins on a
fixed lattice is given by:

\begin{equation}
  \mathcal{H}_{mag} \left(\bm{r},\bm{s} \right) = -\mu_{B}
  \mu_0\sum_{i=0}^{N}g_{i} \bm{s}_{i} \cdot
  \bm{H}_{ext}-\sum_{i,j,i\neq j}^{N} {J} \left(
  r_{ij} \right)\, \bm{s}_{i}\cdot \bm{s}_{j}
  \label{s_hamiltonian}
\end{equation}

with $g_{i}$ the Land\'e factor of the spin $i$, $\mu_0 = 4\pi
\times 10^{-7}$ kg.m.A$^{-2}$.s$^{-2}$ the vacuum permeability, and
$\mu_{B} = 9.275 \times 10^{-24}$ J.T$^{-1}$ the Bohr's magneton.

The first term in the RHS of Eq.~(\ref{s_hamiltonian}), usually
referred to as the Zeeman term, is the interaction energy acquired by
the spins subject to an external magnetic uniform field ${\bm
  H}_{{ext}}$. This field can be constant, or can also vary in
time, for simulations of electronic paramagnetic experiments.

The second term is the exchange interaction.  It is responsible for
the local alignment of neighboring spins.  The dependence of the
Heisenberg coupling constant ${J}$ on the inter-atomic distance is
a well known quantum result that reveals the Pauli exclusion principle
\cite{kittel_quantum_1987}.  This also makes possible a natural
connection to Joule magnetostriction and plays a fundamental role in
the coupling between the spin and lattice degrees of freedom. The
prefactors $g_{i}$ and $g_{j}$, which connect two physical
magnetic moments, are generally absorbed into a redefinition of ${\rm
  J}$.  Details about the parametrization of ${J}$ as a function
of the inter-atomic distance $r_{ij}$ are given in
\ref{app:exchange}.


The exchange interaction mediates a natural coupling between the spin
and lattice degrees of freedom due to the dependence of ${J}\left(
r_{ij} \right)$ on the interatomic distance.  Because ${J}$ is
generally a radial function only, no anisotropic effect can be modeled
in this manner. This eliminates the most interesting and
technologically appealing magnetostriction properties of materials
which are mostly direction dependent and come from the
magneto-crystalline anisotropy energy of materials.  Because the
origin of the magneto-crystalline anisotropy energy is in the
spin-orbit coupling of atoms, it is necessary to take this interaction
into account to perform accurate and realistic spin--lattice
simulations.

Additional terms, responsible for all the quantum aspects of spin are
also of fundamental importance and often need to be considered as
well. This includes both the local and non-local anisotropic
interaction, responsible for the alignment of the spins along
preferred directions, the anti--symmetric pair couplings known as the
Dzyaloshinskii-Moriya's interaction, responsible for weak magnetism of
canted spin spirals, and the long--range dipolar interaction,
responsible for macroscopic magnetic textures in domains.


In this context, issues related to the treatment of spin--orbit
coupling were discussed recently by Perera \emph{et al.} in
Ref.~\cite{perera_reinventing_2016}.  In particular, they showed that
the exchange interaction on its own is not sufficient to simulate
correctly the transfer of energy from the lattice to the spins.  As
these energies are extremely small compared to other electronic
energies, their evaluation via first-principles calculations would
require very high accuracy and thus such a model is difficult to
achieve \cite{victora_theory_1993}.

For this reason Beaujouan \emph{et
  al.}\cite{beaujouan_anisotropic_2012} and later Perera \emph{et
  al.}\cite{perera_reinventing_2016} proposed specific approximations
to simulate this spin-orbit coupling.  Beaujouan used short range
pseudo-dipolar and pseudo-quadrupolar functions, first introduced by
N\'eel \cite{neel_approche_1954} and discussed later by Bruno
\cite{bruno_magnetic_1988}, for the simulation of bulk
magnetostriction and surface anisotropy in cobalt. Forms are given in
\ref{subapp1:soc}.

Once $\mathcal{H}_{mag}$ is formed as a functional of spin
vectors, the time dynamics of the ensemble of spins on a fixed
lattice, labelled by $i$, can be simulated by integrating the set of
coupled EOM.

\subsection{Equations for spin--lattice dynamics}\label{subsec2:dynamics}

For the EOM of the coupled spin--lattice system, we describe the
lattice (nuclear coordinates of atoms) with the classical variables of
molecular dynamics.  Each atom $i$ in the system stores a position
vector $\bm{r}_{i}$ and momentum vector $\bm{p}_{i}$, in
addition to the spin vector $\bm{s}_{i}$.

This allows us to introduce a spin--lattice Hamiltonian

\begin{equation}
  \mathcal{H}_{sl}\left(\bm{r}, {\bm p}, \bm{s} \right) =
  \mathcal{H}_{mag}\left(\bm{r},\bm{s} \right) + \sum_{
    i=1}^{N}\frac{|\bm{p}_{i} |^2}{2m_{i}} + \sum_{{
      i,j}=1}^{N} V\left( r_{ij} \right), \label{sl_hamiltonian} \\
\end{equation}

where, on the first term is the spin Hamiltonian described in
Eq.~\ref{s_hamiltonian}.  The second term is the kinetic energy of the
particles, and the last term is a mechanical potential, e.g.~a pair
potential, binding the atoms together.

In previous implementation of spin--lattice simulations, different
choices have been made for this mechanical potential.  Ma \emph{et
  al.} \cite{ma_large-scale_2008} and Dilina \emph{et al.}
\cite{perera_reinventing_2016} chose potentials derived for large
scale simulations of magnetic $\alpha$-Fe by Dudarev and Derlet
\cite{dudarev_`magneticinteratomic_2005}.  Beaujouan \emph{et al.}
\cite{beaujouan_anisotropic_2012} used an Embedded-Atom Method model
(EAM) derived and parametrized for several metallic elements including
cobalt \cite{thibaudeau_embedded-atom_2008}. These spin-less
potentials are constructed in a way that avoids double-counting the
ferromagnetic contribution to the mechanical energy.  Both of these
potentials correctly reproduce known magneto-elastic properties.
\ref{app:mechpotential} discusses the consequences of the choice of
such potentials.

Once a spin-less mechanical potential is chosen, the EOM can be
derived from the spin--lattice Hamiltonian. Yang \emph{et al.} derived
a generalized formulation of the Poisson bracket for spin--lattice
systems \cite{yang_generalizations_1980}.  With $f\left(t, \bm{r}_{
  i}, \bm{p}_{i}, \bm{s}_{i} \right)$ and $g\left(t,
\bm{r}_{i}, \bm{p}_{i}, \bm{s}_{i} \right)$ two functions
of time, position, momentum and spin:

\begin{equation}
 \lbrace f,g\rbrace = \sum_{i = 1}^{N} \left[ \frac{\partial
     f}{\partial \bm{r}_{i}} \cdot \frac{\partial g}{\partial
     \bm{p}_{i}} - \frac{\partial f}{\partial \bm{p}_{i}}
   \cdot \frac{\partial g}{\partial \bm{r}_{i}} -
   \frac{\bm{s}_{i}}{\hbar} \cdot \left(\frac{\partial f}{\partial
     \bm{s}_{i}} \times \frac{\partial g}{\partial \bm{s}_{i}}
   \right) \right].
\end{equation}

Its application to the spin--lattice Hamiltonian defined by
Eq.~(\ref{sl_hamiltonian}) leads to the following set of EOM for the
spin--lattice system:

\begin{eqnarray}
  \frac{d\bm{r}_{i}}{dt} &=&\lbrace \bm{r}_{i}, \mathcal{H}_{sl} \rbrace
                                                =\frac{\bm{p}_{i}}{m_{i}} \label{r_advance} \\
 \frac{d\bm{p}_{i}}{dt} &=& \lbrace \bm{p}_{i}, \mathcal{H}_{sl} \rbrace \nonumber \\
					    &=& \sum_{j,i\neq j}^{N}\left[-\frac{dV\left(r_{ij}\right)}{dr_{ij}}+
\frac{dJ\left( r_{ij}\right)}{dr_{ij}} \bm{s}_{i} \cdot \bm{s}_{j} \right] \bm{e}_{ij} \label{p_advance}\\
\frac{d\bm{s}_{i}}{dt} &=& \lbrace \bm{s}_{i}, \mathcal{H}_{sl} \rbrace = \bm{\omega}_{i}\times\bm{s}_{i}\label{s_advance}
\end{eqnarray}

where $\bm{e}_{ij}$ is the unit vector along  $\bm{r}_{ij}$.

Eq.~\ref{r_advance} is the standard equation for propagation of
positions in any MD simulation, and Eq.~\ref{s_advance} is an undamped
version of the equation for propagation of the spins (see
Eq.~\ref{LLG}).

The propagation of momenta in Eq.~\ref{p_advance}, includes not only
the mechanical force (derivative of the interatomic potential), but
also a magnetic force varying with the inter-atomic distance and spin
configurations, here including only the exchange interactions.
Extensions to other magnetic interactions are given in
refs.\cite{beaujouan_anisotropic_2012,perera_reinventing_2016}.  It is
also interesting to note that these equations correspond to those
derived from a quantum formalism by Antropov \emph{et al.} in
Ref.~\cite{antropov_spin_1996}.  The spin--lattice dynamics is thus
determined by the numerical integration of this set of coupled
differential equations for each atom $i$ in the system, a total of
$9\times N$ equations for a system of $N$ atoms.

\section{Integration algorithm}\label{sec:algorithms}
 
The set of EOM detailed in Section~\ref{sec:dynamics} define the
coupled spin-lattice dynamics in the microcanonical ensemble. In order
to generate numerical solutions to these equations, they must be
approximated by equivalent discretized equations.  In order to
preserve the geometric properties of the underlying Hamiltonian
equations, the discretized equations must preserve certain fundamental
properties. The most important of these are micro-reversibility in
time, the symplectic character and conservation of the phase space
volume\cite{tuckerman_statistical_2010}.

For notational simplicity, we gather the spin, position and momentum
variables into a single vector $\bm{X}(t)$:

\begin{equation}
 \bm{X}(t) =
\left(
\begin{array}{c}
\bm{r}(t)\\
\bm{p}(t)\\
\bm{s}(t)
\end{array}
\right)
\end{equation}

With this definition in hand, one can write Eqs.~\ref{r_advance},
\ref{p_advance} and \ref{s_advance} formally as a single first-order
differential equation~:

\begin{equation}
 \frac{d\bm{X}(t)}{dt} = \hat{L} \bm{X}(t)
\end{equation}

where $\hat{L}$ is a Liouville operator of the spin--lattice
system. This can be viewed itself as a sum of Liouville operators
$\hat{L}_{r}$, $\hat{L}_{p}$ and $\hat{L}_{s}$, acting on
each variable separately

\begin{eqnarray}
  \hat{L} &=& \sum_{i =1}^N \left( \frac{d\bm{r}_{i}}{dt} \cdot
  \frac{\partial }{\partial \bm{r}_{i}} + \frac{d\bm{p}_{
      i}}{dt} \cdot \frac{\partial }{\partial \bm{p}_{i}}+
  \frac{d\bm{s}_{i}}{dt} \cdot \frac{\partial }{\partial
    \bm{s}_{i}}\right) \nonumber\\ &\equiv& \hat{L}_{r} +
  \hat{L}_{p} + \hat{L}_{s} \label{Liouville}
\end{eqnarray}

Integration of the vector $\bm{X}(t)$ from a time $t$ to $t+\Delta t$
can be interpreted formally as the application of the exponential of
the Liouville operator $\hat{L}$ on the vector ${\bm X}(t)$

\begin{equation}
  \bm{X}(t+\Delta t) = e^{ \left( \hat{L}_{r} + \hat{L}_{p} +
    \hat{L}_{s}\right) \Delta t }\bm{X}(t)
\end{equation}

However, the individual operators that form $\hat{L}$ do not
necessarily commute with each other.  A convenient way to generate an
algorithm that deals with these non-commuting operations is to
recognize this system as a Magnus expansion \cite{blanes_magnus_2009}
and to apply a Suzuki--Trotter (ST) decomposition to the exponential
of $\hat{L}$

\begin{equation}
  e^{\left(\hat{L}_{p} + \hat{L}_{r} + \hat{L}_{
      s}\right)\Delta t} = e^{\hat{L}_{p}\frac{\Delta t}{2}}
  ~e^{\hat{L}_{s}\frac{\Delta t}{2}} ~e^{\hat{L}_{r}\Delta t}
  ~e^{\hat{L}_{s}\frac{\Delta t}{2}} ~e^{\hat{L}_{
      p}\frac{\Delta t}{2}} +\mathcal{O}\left( \Delta t^3
  \right) \label{general_ST_decomposition}
\end{equation}

which is accurate to $\mathcal{O}\left( \Delta t^3 \right)$ in the timestep.

Many other similar decompositions can be obtained by permuting the
three operators.  They are all formally equivalent in that they are
all accurate to $\mathcal{O}\left( \Delta t^3 \right)$. However, they
do not produce the same overall numerical accuracy
\cite{beaujouan_anisotropic_2012}. The decomposition of
Eq.~\ref{general_ST_decomposition} is particularly effective because
the timestep necessary for integration of the spin system is generally
an order of magnitude smaller than the one commonly used for classical
molecular dynamics of atomic systems.  As accuracy is our main
objective, we do not put the spin operator in the middle of the ST
decomposition; individual updates are thus performed with 2x smaller
timesteps.  In addition, if there is no spin in the system, then
$\hat{L}_s=\hat{0}$ and $\exp(\hat{L}_s\Delta t)=\hat{1}$ and the
resulting combination produces the well known Verlet leapfrog
algorithm \cite{frenkel_understanding_2002}, in the form of
velocity-Verlet integration scheme.


Finally, putting $\hat{L}_{r}$ at the center of the ST
decomposition ensures that the mechanical forces need only be
calculated once per timestep, which greatly simplifies the practical
implementation of the method, particularly for the parallel algorithm
(see Section~\ref{sec:parallel}).

Because the equations of motion for the spins are first-order
differential equations directly depending on the neighboring spin
orientations (due to the strong short--range interactions), the
individual spin rotations in 3-dimensional space do not commute with
each other.  Therefore, propagating a spin before one of its
neighbors, or the opposite way around will give different results.
Thus the global spin operator $\hat{L}_{s}$ must be further
decomposed into a sum of operators $\hat{L}_{s_i}$, each one being
the time integration operator of a given spin $i$.  One has the
following ST decomposition for the exponential of $\hat{L}_s$:

\begin{eqnarray}
  e^{\hat{L}_{s}\frac{\Delta t}{2}} &=& e^{\hat{L}_{
      s_1}\frac{\Delta t}{4}} \dots~e^{\hat{L}_{s_{N}}\frac{\Delta
      t}{2}}~\cdots ~e^{\hat{L}_{s_1}\frac{\Delta t}{4}}
  +\mathcal{O}\left( \Delta t^3 \right) \label{spin_ST_decomposition}
  \nonumber \\ &=& \prod_{{ i}=1}^{N} e^{\hat{L}_{
      s_{i}}\frac{\Delta t}{4}} \prod_{{ i=N}}^{1}
  e^{\hat{L}_{s_{i}} \frac{\Delta t}{4}} +\mathcal{O}\left( \Delta
  t^3 \right)
\end{eqnarray}

In practice, this rule means that the magnetic force $\bm{\omega}_{
  i}$ acting on the spin $i$ has to be computed immediately before
updating the value of the spin $i$.  Indeed, the updated values of the
neighboring spins of the spin $i$ have to be taken into account in the
calculation of $\bm{\omega}_{i}$ \cite{omelyan_algorithm_2001}.

Also, as discussed in Section~\ref{sec:dynamics}, the spin--spin
interaction coefficients depend now on the inter--atomic distances.
Because the atoms positions are updated between the two spin updates,
the inter--atomic distances change, and thus the coefficients have to
be computed again.  This is one of the main disadvantages of the
global decomposition detailed by Eq.~\ref{general_ST_decomposition},
compared to a decomposition for which the propagation operator of the
spins $e^{\hat{L}_{s}\Delta t/2}$ would be the central operator.
However, as already mentioned, the applied decomposition was mainly
chosen for accuracy reasons, and it also allows us to compute the
mechanical force once per timestep.

The fact that the precession vector $\bm{\omega}_{i}$ has to be
computed before updating the spin $i$, and requires knowledge of the
configuration of all neighboring spins requires that spin updates be
performed in a sequential manner.  This prevents the straightforward
application of parallel algorithms that rely on concurrent updates.
In Section~\ref{sec:parallel}, we present a synchronous sublattice or
sectoring algorithm that does not suffer from these limitations. It
enables efficient simulation of spin-lattice systems with non-uniform
or disordered spatial structure on large parallel computers, while
preserving the properties of the ST decomposition.

Eqs.~\ref{LLG} and \ref{s_advance} exhibit a model of precession that
preserves the norm of each individual spin over time. Consequently,
its corresponding numerical time evolution operator must preserve this property
to a given accuracy as well.  From geometrical considerations, Omelyan
\emph{et al.} \cite{omelyan_algorithm_2001} derived the following
expression for the single-spin propagation operators~:

\begin{widetext}
\begin{equation}
  \bm{s}_{i} (t+\Delta t) = e^{\hat{L}_{\bm{s}_{i}}\Delta t
  }\bm{s}_{i} (t) = \Bigg\{ \bm{s}_{i}(t) + \Delta
  t\left(\bm{\omega}_{i}(t) \times \bm{s}_{i}(t) \right)
  +\frac{\Delta t^2}{4} \Big[ 2\,\bm{\omega}_{i}(t)
    \left(\bm{\omega}_{i}(t) \cdot \bm{s}_{i}(t)\right) -
    \|\bm{\omega}_{i}(t)\|^2\bm{s}_{i}(t) \Big] \Bigg\}
  \Biggm/ \left( 1+\frac{\Delta t^2}{4}\|{\bm\omega}_{
    i}(t)\|^2\right) + \mathcal{O} \left( \Delta t^3 \right)
\end{equation}
\end{widetext}

This expression uses low-order approximants for trigonometric
functions that exactly satisfy $\cos^2(\|\bm{\omega}_i\|\Delta
t)+\sin^2(\|\bm{\omega}_i\|\Delta t)=1$, unlike the usual Pad\'e
approximants. As a result, the single-spin propagator preserves the
same $\mathcal{O} \left( \Delta t^3 \right)$ accuracy as the rest of
the discretization scheme.  Moreover, the number of numerical
operations needed to evaluate this time increment is lower than the
corresponding calls to the trigonometric functions, which
significantly speeds up the procedure and eliminates the need of a
norm rescaling that usually breaks the symplectic character and time
reversal properties \cite{dullweber_symplectic_1997}.

\begin{figure}[ht]
\centering
\includegraphics[width=0.95\columnwidth]{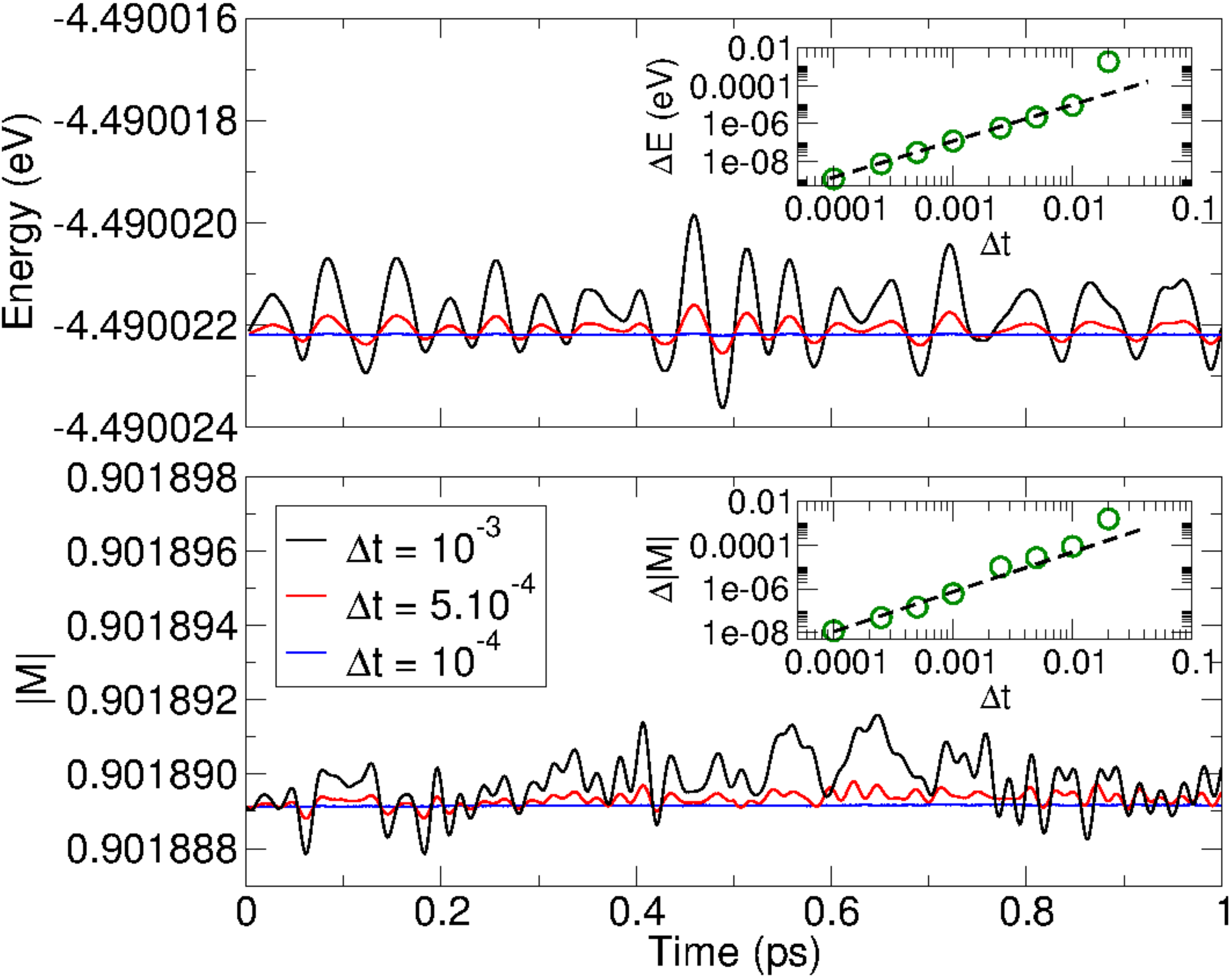}
\caption{NVE SD--MD simulations of a block of 500 fcc cobalt atoms
  with periodic boundary conditions, with initial conditions
  corresponding to a temperature of 300~K.  The graphs plot the energy
  per atom and the norm of the total magnetization $|{\rm \bm{M}}|$
  for timesteps sizes of $10^{-3}$~ps (black), $5\cdot 10^{-4}$~ps
  (red), and $10^{-4}$~ps (blue).  The two insets plot the mean
  relative absolute deviations of the energy $\Delta {E}$ and
  magnetization $\Delta |{\rm M}|$ according to
  Eq.~\ref{relative_deviation} as a function of the timestep size
  $\Delta t$. The black dashed lines in the insets represent $
  \mathcal{O} \left( \Delta t^2 \right)$ scaling
  behavior.} \label{fig:timestep_accuracy}
\end{figure}

Once the spin vector is computed, both the position and momentum are
updated via the Verlet method already developed in LAMMPS
\cite{plimpton_lammps-large-scale_2007}, and fully coupled
spin--lattice simulations can now be performed.  To evaluate the
numerical efficiency of this algorithm, the average total energy and
the norm of the magnetization must be measured as a function of the
timestep. \cite{omelyan_algorithm_2001}.  Representative numerical
simulations were constructed with an fcc crystal of 500 cobalt
atoms. The individual magnetizations were assigned random initial
orientations selected from an equilibrium distribution corresponding
to a temperature of 300 K, consistent with the Curie-Langevin law. The
system was then evolved for 1~ps using the NVE time integration
algorithm described above using three different timestep sizes.  The
total energy and total magnetization of the system are shown in
Fig.~\ref{fig:timestep_accuracy}.

Because the system is closed, it cannot adiabatically exchange energy
with a surrounding reservoir and both quantities are very well
conserved.  Typical spin dynamics simulations use a timestep of
$\Delta t = 10^{-4}$~ps.  The current algorithm remains reasonably
accurate up to a timestep size of $\Delta t = 10^{-2}$~ps.  This
demonstrates the strong stability of the integration algorithm.

The influence of the timestep on the fluctuations in these quantities
can be examined more precisely by calculating the mean relative
absolute deviations averaged over the trajectory.  For example, in the
case of total energy per atom

\begin{equation}
  \Delta {E} (\Delta t) = \frac{1}{N_{step}}
  \sum_{k=1}^{N_{step}} \left\lvert \frac{{E}_{k}(\Delta t) -
    \langle {E} \rangle (\Delta t) }{\langle {E} \rangle
    (\Delta t)} \right\rvert,
  \label{relative_deviation}
\end{equation}

where $E_{k}$ is the total energy per atom at timestep $k$ and 
$\langle {E} \rangle$ is the time average of the total energy per atom.


The same formula can be applied to the norm of the total
magnetization.  The dependence of both of these quantities on the
timestep size is shown in the two insets of
Fig.~\ref{fig:timestep_accuracy}.  The deviations increase as
$\mathcal{O} \left( \Delta t^2 \right)$ . This behavior is consistent
with the $\mathcal{O} \left( \Delta t^3 \right)$ accuracy of the ST
decomposition, which describes propagation over a single step.  For
$N$ consecutive steps, the accuracy of the decomposition algorithm is
given by $\mathcal{O} \left( N \Delta t^3 \right)$, with
$N=t_{total}/\Delta t$.  Thus, the accuracy for a fixed time interval
is $\mathcal{O} \left( \Delta t^2 \right)$, as seen in
Fig.~\ref{fig:timestep_accuracy}.  For timestep size greater than
$\Delta t = 10^{-2}$~ps, the scaling law starts to break down,
indicating the the limit of numerical stability is being approached.

\section{Spin--lattice dynamics in a canonical ensemble (NVT)}\label{sec:temperature}

Now we connect the spin--lattice system to an external energy
reservoir and stabilize the exchanges of energy with a given
thermostat. In MD, many approaches have been already explored
\cite{andersen_molecular_1980,nose_unified_1984,hoover_canonical_1985}
and we focus on a Langevin strategy \cite{feller_constant_1995}.  This
is a widely used thermostat for atomic systems as well as
coarse-grained models, e.g.~dissipative particle dynamics
\cite{hoogerbrugge_simulating_1992} (see \cite{groot_dissipative_1997}
for a review).  In the Langevin approach, both random forces and
damping terms are used in accord with the fluctuation-dissipation
theorem. However, the presence of damping terms will lead to a 
contraction of the phase space volume, producing
non-Hamiltonian dynamics. Yet, even in these cases, it has been shown
that if some phase-space related measures are not preserved,
pathological situations can be observed, and lead to incorrect
distributions \cite{tuckerman_non-hamiltonian_2001}.

\subsection{Connecting spins to a random bath}\label{subsec1:temperature}

Néel first \cite{neel_influence_1949} and Brown later
\cite{brown_thermal_1963} demonstrated that a sufficiently fine
ferromagnetic particle consists of a single magnetic structure in
which thermal agitation causes continual changes in the orientation of
the moment.  Such thermal fluctuations in the magnetization are well
described by following the Langevin approach
\cite{antropov_ab_1995,antropov_spin_1996,garcia-palacios_langevin-dynamics_1998}.
In this approach, the spin system is connected to a single thermal
bath modeled by an infinite number of degrees of freedom.  The
properties of such bath are given by $\bm{\eta}$, a random vector,
whose components follow a Gaussian probability law given by the first
and second moment

\begin{eqnarray}
 \langle {\bm\eta}(t) \rangle &=& {\bm 0} \nonumber \\
 \langle \eta_{\alpha}(t) \eta_{\beta}(t') \rangle &=& 2 D_S \delta_{\alpha\beta}\delta (t-t') \label{white_noise}
\end{eqnarray}

where $\alpha$ and $\beta$ are the vector components.

In order to preserve the norm of each individual spin, this random
fluctuation is added in a multiplicative manner to Eq.~\ref{LLG} via a
random torque \cite{mayergoyz_nonlinear_2009}, that leads to the
following stochastic Landau-Lifshitz-Gilbert equation~:

\begin{equation}
 \frac{d \bm{s}_{i}}{dt} = \frac{1}{\left(1+\lambda^2 \right)} \left( \left(\bm{\omega}_{i} +\bm{\eta} \right) \times \bm{s}_{i} + \lambda\, \bm{s}_{i}\times\left( \bm{\omega}_{i} \times\bm{s}_{i} \right) \right). \label{sLLG}
\end{equation}

From Eq.~\ref{sLLG}, one can derive a Fokker-Planck (FP) equation
\cite{garcia-palacios_langevin-dynamics_1998}. The resolution at
equilibrium of the FP equation allows a fluctuation--dissipation
relation to be derived that assigns a proportion of the amplitude of
the noise $D_S$ to the given external thermostat
$T$~\cite{mayergoyz_nonlinear_2009}, such that

\begin{equation}
  D_S = \frac{2 \pi \lambda k_B {T}
  }{\hbar}, \label{spin_fluctuation_dissipation}
\end{equation}

with $k_B$ = the Boltzmann constant.

A subtle detail about stochastic differential equations is a proper
choice of a stochastic prescription that allows an evaluation of the
ill-defined random vector~:

\begin{equation}
  \bm{\mathcal{I}}(\Delta t) ~=~ \int_{t}^{t+\Delta t} \bm{\eta}(t') \times \bm{s}_{i}(t') dt',
\end{equation}

by defining a point within the interval $[t, t+\Delta t]$ at which
this integral is approximated consistently.  Numerous papers have
discussed various choices of the prescription for stochastic
magnetization dynamics performed with the stochastic LLG equation
\cite{garcia-palacios_langevin-dynamics_1998,daquino_midpoint_2006,aron_magnetization_2014}.
We use the mid--point Stratonovich approach that preserves the norm of
individual spins and has time micro-reversibility.

\subsection{Measuring lattice and spin temperatures}\label{subsec3:temperature}

In non-equilibrium spin-lattice systems, it is convenient to use a
single thermostat to control the flow of entropy production between
the system and a thermal reservoir \cite{rapaport_art_2004}.  Such a
thermostat has to be related to a microcanonical quantity defined in a
statistical ensemble that measures the temperature.  Because the
temperature of an equilibrium system is calculated from the mean
kinetic energy of its particles, the transient kinetic energy for
spins is not an obvious quantity.  An immediate consequence is how to
measure the temperature $T_{S}$ that controls the thermostatting
of the magnetic degrees of freedom \cite{antropov_spin_1996}.

At equilibrium, the instantaneous lattice temperature is usually
defined as the kinetic temperature of the atoms:

\begin{equation}
  T_{L} = \frac{2}{3\, N\, k_B} \sum_{i=1}^{N} \frac{|\bm{p}_{
      i}|^2}{2 m_{i}}. \label{lattice_temperature}
\end{equation}

This expression of the kinetic temperature is known to rely on some
approximations, such as the equivalence of ensembles and the ergodicity
assumption \cite{rugh_dynamical_1997}.
Other dynamical approaches for measuring the temperature in Hamiltonian systems
within a microcanonical ensemble have been derived, and proved to be accurate
\cite{rugh_geometric_1998,jepps_microscopic_2000}.
However, we will assume that the kinetic expression given by eq.~\ref{lattice_temperature} is sufficient for the purpose of this work.

Transcripting Rugh's the geometrical approach to spin systems
\cite{rugh_dynamical_1997}, and when the thermodynamic limit is considered, Nurdin \emph{et al.} \cite{nurdin_dynamical_2000} give a spin temperature as

\begin{equation}
  T_{S} = \frac{\hbar}{2 k_B} \frac{\sum_{i=1}^{N} |\bm{s}_{i}
    \times \bm{\omega_{i}}|^2}{\sum_{i=1}^{N} \bm{s}_{i} \cdot
    {\bm\omega}_{i}} \label{spin_temperature}
\end{equation}

Another approach was later derived by Ma \emph{et al.} \cite{ma_temperature_2010}. 
Relying on the fluctuation-dissipation theorem, it gave an analog definition of the temperature of a spin ensemble.

In refs.~\cite{beaujouan_anisotropic_2012,perera_reinventing_2016},
the definition given by eq.~\ref{spin_temperature} has proven to be efficient for the thermalization of the spin subsystem and its relaxation toward thermal equilibrium
during spin--lattice simulations. Besides, this definition of $T_{S}$ has
the same domain of validity as the kinetic temperature expressions for
$T_{L}$ defined above.
Thus, we chose to use the definition of the spin temperature defined by Nurdin
\emph{et al.}.

\subsection{Thermalizing the spin--lattice system}\label{subsec2:temperature}

For the simulation of relaxation processes, the connecton of the
lattice system to a thermal bath is also performed using the Langevin
approach.  Eq.~\ref{s_advance} is replaced by Eq.~\ref{sLLG}, and a
new random force and a damping term are added to Eq.~\ref{p_advance}.
This yields the following equations that model stochastic magnetic
molecular dynamics:

\begin{eqnarray}
 \frac{d\bm{r}_{i}}{dt} &=& \frac{\bm{p}_{i}}{m_{
     i}} \label{r_advance_T} \\ \frac{d\bm{p}_{i}}{dt} &=&
 \sum_{j,i\neq j}^{N}\left[-\frac{dV\left(r_{
       ij}\right)}{dr_{ij}}+ \frac{dJ\left( r_{
       ij}\right)}{dr_{ij}} \bm{s}_{i} \cdot \bm{s}_{j}
   \right] \bm{e}_{ij} \nonumber \\ ~&~& -\frac{\gamma_L}{m_{
     i}} \bm{p}_{i}
 +\bm{f}(t) \label{p_advance_T}\\ \frac{d\bm{s}_{i}}{dt} &=&
 \frac{1}{\left(1+\lambda^2 \right)}\left(\left(\bm{\omega}_{i}
 +\bm{\eta}(t)\right) \times \bm{s}_{i} + \lambda\, \bm{s}_{
   i}\times\left( \bm{\omega}_{i} \times\bm{s}_{i} \right)
 \right) \label{s_advance_T}
\end{eqnarray}
In Eq.~\ref{p_advance_T}, $\gamma_L$ is a damping parameter, and $\bm{f}$ its corresponding fluctuating force drawn from a Gaussian distribution with
\begin{eqnarray}
 \langle {\bm f}(t) \rangle &=& {\bm 0} \\
 \langle f_{\alpha}(t) f_{\beta}(t') \rangle &=& D_L \, \delta_{\alpha\beta}\, \delta(t-t')
\end{eqnarray}

where $\alpha$ and $\beta$ are coordinates, and $D_L$ is the amplitude
of the random variables.  $D_L$ can be parametrized according to the
fluctuation--dissipation relation, and then depends on the temperature
of the thermal bath coupled to the lattice, and on the damping
coefficient $\gamma_L$ according to the Einstein relationship
\cite{tuckerman_statistical_2010}.  The probability distribution of
the noise vector $\bm{\eta}$ follows Eqs.~\ref{white_noise} and
\ref{spin_fluctuation_dissipation} respectively.

Extensions to more than two out-of-equilibrium dynamics (here referring to 
spin and lattice dynamics) are possible and several
authors linked the damping coefficients to spin--electron and
lattice--electron relaxation processes
\cite{ma_spin-lattice-electron_2012}.  This allowed using the model
presented above to simulate spin--lattice--electron relaxation
processes that occur in ultrafast magnetic switching experiments.

However there are concerns about the choice of the noise correlation
functions.  In this study, we decided to remain within the framework
of the Markov hypothesis, and focus on uncorrelated white-noise only.
When the characteristic timescales of the dynamics reach values as
small as those of the simulated relaxation processes, the Markov
hypothesis may break down and a colored-noise, such as an
Ornstein–Uhlenbeck process, becomes a better approximation of the
exchange of causal information between the system and its reservoir.
Recent studies have focused on evaluating the influence of such memory
effects on the magnetization dynamics
\cite{atxitia_ultrafast_2009,bose_correlation_2010,tranchida_closing_2016,tranchida_colored-noise_2016},
and suggest non-trivial magnetization dynamics beyond the second order
cumulant of the spin variables. These points will be addressed in a
separate study.

In order to evaluate the efficiency of the two-thermostat model
presented here, two different simulations were performed.  A cell of
500 cobalt atoms on an fcc lattice, coupled by three interactions (the
magnetic exchange interaction, a spin-orbit coupling, and a mechanical
EAM potential
\cite{thibaudeau_thermostatting_2012,pun_embedded-atom_2012}) is
considered.

For the first simulation, a Langevin thermostat was applied only to
the spins, according to Eq.~\ref{s_advance_T}.  The simulation started
from a fixed lattice equilibrium configuration, corresponding to
$T_{L}=0$K and an initial spin configuration sampled from a
$T_{S}=300$K magnetic equilibrium state.

\begin{figure}[ht]
\centering
\includegraphics[width=0.95\columnwidth]{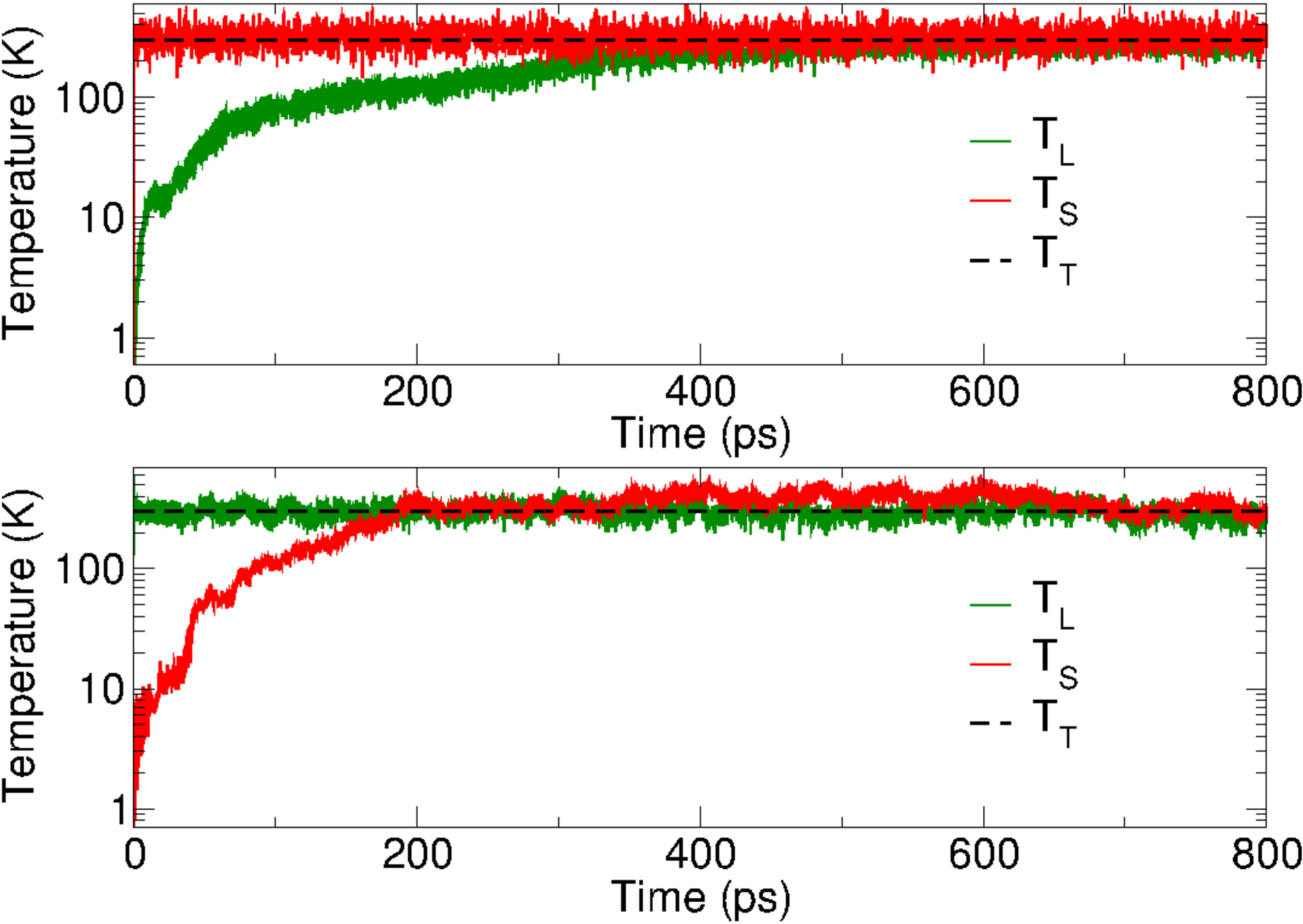}
\caption{Measure of the spin T$_{S}$ and lattice T$_{L}$
  temperatures during two NVT simulations of 500 cobalt atoms.  For
  both graphs, the black dashed lines plot the target temperature
  T$_{T}$, and the green and red curves plot the lattice and spin
  temperatures, T$_{L}$ and T$_{S}$, respectively.  In the
  upper graph, the Langevin thermostat was applied to the spins only,
  according to Eq.~\ref{s_advance_T}, with a target T$_{T}$=300K,
  and the transverse magnetic damping set to $\lambda = 0.01$.  In the
  lower graph, the thermostat was applied only to the atomic motion,
  according to Eq.~\ref{p_advance_T} with T$_{T}$=300K and a
  lattice damping parameter of $\gamma_L=10$ s$^{-1}$.}
\label{fig:spinlatt_relaxation}
\end{figure}

The second simulation was exactly the opposite: the Langevin
thermostat was applied only to the motion of the lattice atoms,
according to Eq.~\ref{p_advance_T}.  The simulation started in a
configuration with all the spins aligned along their effective fields,
which corresponds to $T_{S}=0$K, and with atom velocities sampled
from a $T_{L}=300$K equilibrium lattice state.
Fig.~\ref{fig:spinlatt_relaxation} plots the time evolution of $T_{
  L}$ and $T_{S}$ for the two simulations.

In each case, the thermostatted degrees of freedom stayed at the
target temperature for the duration of the simulation.  And the
non-thermostated degrees of freedom relaxed from their initial
temperature to the thermostatted temperature due to the spin--lattice
coupling of the dynamics.  In both cases the relaxation time was a few
100 ps, which is consistent with the spin--lattice coupling time shown
in Fig.~\ref{fig:3T}.  This illustrates how proper choice of damping
coefficients can produce physically correct responses for magnetic
exchange, spin-orbit, and mechanical coupling interactions.

\begin{figure}[ht]
\centering
\includegraphics[width=0.95\columnwidth]{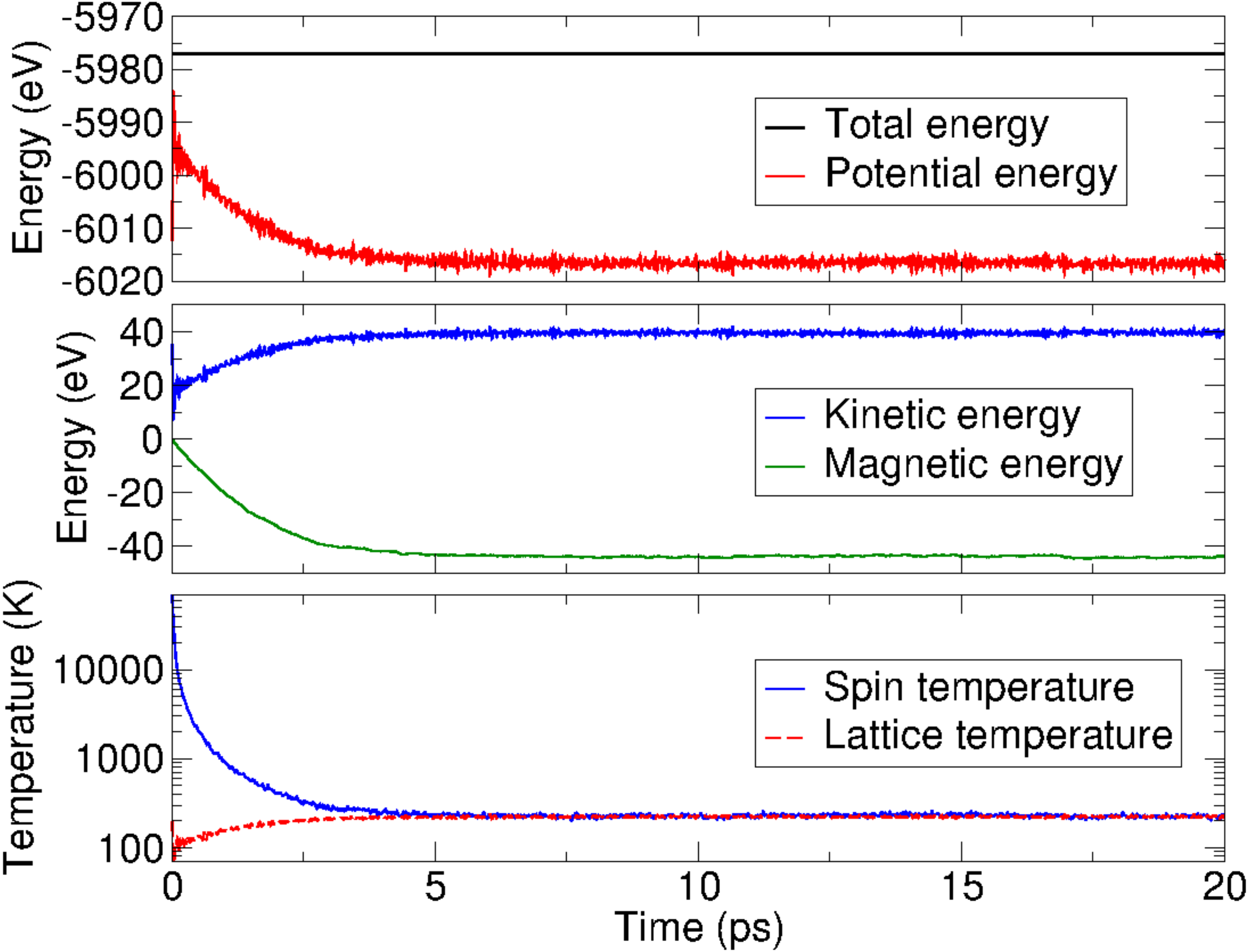}
\caption{NVE simulation of a periodic block of 1372 cobalt atoms.
  Spin directions are initially random.  Initial particle velocities
  were drawn from a Maxwell distribution at 200K.  The upper graph
  plots 3 contributions to the total energy which remained constant:
  kinetic energy of the atoms and potential energies of the atoms and
  spins (magnetic energy).  The lower graph plots the atomic (lattice)
  and spin temperatures, corresponding to
  Eqs.~\ref{lattice_temperature} and \ref{spin_temperature}
  respectively, which equilibrated to the same value.}
\label{fig:energies}
\end{figure}

To better understand how the magnetic and mechanical energy couple to
each other, a larger NVE simulation was performed with 1372 cobalt
atoms initially on an fcc lattice.  The initial spin configuration was
random so that the spin system is in a paramagnetic state at a very
high temperature.  The atomic velocities (lattice temperature) were
initialized to 200K.  The potential energy of the atoms was initially
at a minimum due to the perfect lattice configuration.  As the
simulation evolved energy re-partitioned between these 3 components
(kinetic energy, potential energies of atoms and spins) but the total
energy remained constant with no perceptible fluctuations (at this
scale) as seen in the upper graph of Fig.~\ref{fig:energies}.  The
lower graph of Fig.~\ref{fig:energies} plots the evovling spin and
lattice temperatures, which equilibrated to the same value, indicating
the efficacy of the spin--lattice coupling.

\section{Parallel implementation of the spin--lattice algorithm}\label{sec:parallel}

\subsection{Synchronous sub-lattice algoritm}\label{subsec1:parallel}

To exploit the power of parallel processing, we developed a
spatially-decomposed version of the serial SD--MD algorithm described
in Section~\ref{sec:algorithms}, analogous to that used for molecular
dynamics simulations in LAMMPS.  Like the serial algorithm, it is
essential that the parallel algorithm accurately conserve energy and
the norm of the total magnetization.  As explained in
Section~\ref{sec:algorithms}, this requires that the spin propagation
operators be applied in a manner that preserves symplecticity.  This
in turn requires that the magnetic force ${\boldsymbol \omega}_i$
acting on each spin ${\boldsymbol s}_i$ be calculated when each spin
is updated.  Due to short-range spin-spin interactions, the
computation of this force depends on the spin orientations of the
neighbors of atom $i$.  In other words, before updating each spin, the
current value of its neighboring spins must be known.

The parallel issue arises when two neighbor spins ${\boldsymbol s}_i$
and ${\boldsymbol s}_j$ (and their associated atoms) reside on
different processors.  How do we ensure that when the second spin is
updated, it uses the previously updated value of the first spin?  This
clearly requires some inter-processor communication of spin
information during the update operation, but the cost of the
communication needs to be minimized to ensure an efficient algorithm.
Likewise we must ensure two neighbor spins on different processors
are not updated simultaneously (one of them with outdated information).
Otherwise the accuracy of the integration scheme will be degraded..

Inspired by the lattice-dependent ST decomposition proposed by Krech
\emph{et al.}~\cite{krech_fast_1998}, Ma \emph{et al.}
\cite{ma_parallel_2009} developed a multithreading algorithm for spin
dynamics that partitions the single-spin evolution operators into
groups whose member spins do not interact.  This allows all spins in different
groups to be updated concurrently by separate execution threads.
The multi-threading implementation delivers good
speedup on a single multicore CPU. 
As currently implemented, 
distributed-memory parallel execution is not supported, although
the method could be extended to use spatial 
decomposition parallel algorithms.
The biggest limitation of the method is the lack of generality.
Partitioning into non-interacting groups is achieved using a lattice-coloring or
checkerboarding scheme that can only be applied to perfectly regular
lattices in which each spin interacts with a fixed stencil of
interacting neighbor spins. Moreover, as the number of neighbors in the 
interaction stencil increases, the number of groups required to achieve a
correct ST decomposition also increases, adding to the complexity of the algorithm.

We have implemented a different method called sectoring 
which is based on the synchronous
sub-lattice algorithm \cite{shim_semirigorous_2005,
  plimpton_crossing_2009} used to achieve spatial
parallelism in kinetic Monte Carlo simulations. In contrast to 
checkerboarding, sectoring is very general, 
requiring only that spin-spin interactions 
vanish beyond a finite cutoff distance. As long as this requirement is met,
sectoring can be used for dynamic simulations of arbitrary systems of atomic spins, 
including perfectly regular lattices, thermally vibrating lattices, spatially disordered 
systems, and even systems undergoing diffusive dynamics in which the 
set of neighbors of each spin can change over time.

The sectoring method can be thought of as an extension to the
spatial-decomposition MD algorithm used in LAMMPS and other MD codes,
where the system is partitioned into subdomains, one per processor.
Each processor owns and time-integrates the atoms in its subdomain.
It also stores information about nearby ghost atoms, up to a
cutoff distance away, which are owned by neighboring processors.  This
information is acquired by inter-processor communication, when needed.

The sectoring idea is to further divide each subdomain into smaller
regions by bisecting the subdomain once in each physical dimension (4
sectors in 2d, 8 in 3d).  If the spatial extent of all sectors in any
dimension is larger than the interaction cutoff distance, then spins in
the same sector on two different processors do not interact.  The
processors can thus concurrently update all the spins in one sector
without the need to communicate spin information to/from other
processors, while still adhering to the ST decomposition of
Eq. \ref{spin_ST_decomposition}. Formally, we can rewrite the ST
decomposition as
\begin{equation}
  e^{\hat{L}_{s}\frac{\Delta t}{2}} = \prod_{{ k}=1}^{K} \left \{
					     \prod_{{ j}=1}^{N_k} e^{\hat{L}_{s_{j}}\frac{\Delta t}{4}}
					     \right \} \prod_{{ k}={K}}^1 \left \{
					     \prod_{{ j=N_k}}^{1} e^{\hat{L}_{s_{j}} \frac{\Delta t}{4}}
					     \right \} +\mathcal{O}\left( \Delta t^3 \right), \label{spin_ST_sectoring}
\end{equation}
where ${K}$ is the number of sectors, ${N_k}$ is the number of
spins in sector ${k}$, and ${s_{j}}$ is the ${j}$th spin
in the sector.

 \begin{figure}[ht]
\centering
\begin{tikzpicture}
\draw[line width=2pt] (0,0) rectangle (2,2);
\draw[fill=gray, line width=1pt] (0,0) rectangle (1,1);
\node (a) at (0.5, 0.5) {1};
\draw[line width=1pt] (0,1) rectangle (0,2);
\node (a) at (0.5, 1.5) {3};
\draw[line width=1pt] (1,0) rectangle (2,0);
\node (a) at (1.5, 0.5) {2};
\draw[line width=1pt] (1,1) rectangle (2,2);
\node (a) at (1.5, 1.5) {4};
\draw[line width=2pt] (0,2) rectangle (2,4);
\draw[fill=gray, line width=1pt] (0,2) rectangle (1,3);
\node (a) at (0.5, 2.5) {1};
\draw[line width=1pt] (0,3) rectangle (1,4);
\node (a) at (0.5, 3.5) {3};
\draw[line width=1pt] (1,2) rectangle (2,3);
\node (a) at (1.5, 2.5) {2};
\draw[line width=1pt] (1,3) rectangle (2,4);
\node (a) at (1.5, 3.5) {4};
\draw[line width=2pt] (2,0) rectangle (4,2);
\draw[fill=gray, line width=1pt] (2,0) rectangle (3,1);
\node (a) at (2.5, 0.5) {1};
\draw[line width=1pt] (3,0) rectangle (4,1);
\node (a) at (2.5, 1.5) {3};
\draw[line width=1pt] (2,1) rectangle (3,2);
\node (a) at (3.5, 0.5) {2};
\draw[line width=1pt] (3,1) rectangle (4,2);
\node (a) at (3.5, 1.5) {4};
\draw[line width=2pt] (2,2) rectangle (4,4);
\draw[fill=gray, line width=1pt] (2,2) rectangle (3,3);
\node (a) at (2.5, 2.5) {1};
\draw[line width=1pt] (3,2) rectangle (4,3);
\node (a) at (2.5, 3.5) {3};
\draw[line width=1pt] (2,3) rectangle (3,4);
\node (a) at (3.5, 2.5) {2};
\draw[line width=1pt] (3,3) rectangle (4,4);
\node (a) at (3.5, 3.5) {4};
\end{tikzpicture}
\caption{Schematic of the SD--MD sectoring algorithm for a
  two--dimensional spin--lattice model.  The system is divided into
  four subdomains (bold squares), each owned by a different
  processor.  Each subdomain is further divided into four sectors
  (numbered squares).  Spins in sector 1 (shaded) for two different
  processors do not interact and can thus be updated concurrently
  without interprocessor communication.} \label{fig:sectoring}
\end{figure}
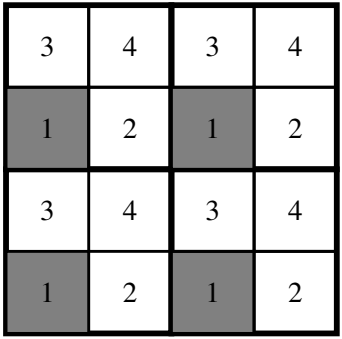

Fig.~\ref{fig:sectoring} illustrates this idea for a two-dimensional
system running on 4 processors, with each processor subdomain further
divided into 4 sectors.  Spins in sector 1 of a particular processor
do not interact with spins in sector 1 of any other processor.  Thus
all the processors can update their sector 1 spins at the same time.
After the sector 1 updates, new spin values must be communicated
between processors before sector 2 spins can be updated.

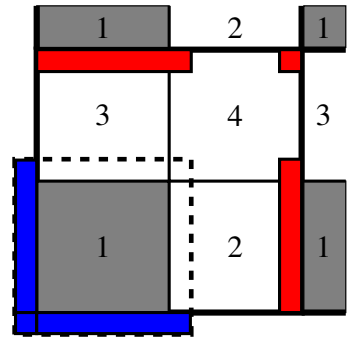
\begin{figure}[ht]
\centering
\begin{tikzpicture}
\draw[line width=2pt] (0,0) rectangle (3,3);

\draw[fill=gray, line width=1pt] (0,0) rectangle (1.5,1.5);
\node (a) at (0.75, 0.75) {1};
\draw[line width=1pt] (0,1.5) rectangle (0,3);
\node (b) at (0.75, 2.25) {3};
\draw[line width=1pt] (1.5,0) rectangle (3,0);
\node (c) at (2.25, 0.75) {2};
\draw[line width=1pt] (1.5,1.5) rectangle (3,3);
\node (d) at (2.25, 2.25) {4};
\draw[style=dashed,line width=1.5pt] (-0.25,-0.25) rectangle (1.75,1.75);

\draw[fill=blue, line width=1.0pt] (-0.24,0.0) rectangle (0.0,1.74);
\draw[fill=blue, line width=1.0pt] (-0.24,-0.24) rectangle (0.0,0.0);
\draw[fill=blue, line width=1.0pt] (0.0,-0.24) rectangle (1.74,0.0);
\draw[fill=red, line width=1.0pt] (2.75,0.0) rectangle (3.0,1.75);
\draw[fill=red, line width=1.0pt] (2.75,2.75) rectangle (3.0,3.0);
\draw[fill=red, line width=1.0pt] (0.0,2.75) rectangle (1.75,3.0);

\draw[fill=gray] (0,3.025) rectangle (1.5,3.5);
\draw[line width=2.0pt] (0,3) -- (0,3.5);
\node (a2) at (0.75, 3.25) {1};
\draw[line width=1.0pt] (1.5,3) -- (1.5,3.5);
\node (b2) at (2.25, 3.25) {2};
\draw[line width=2.0pt] (3.0,3.0) -- (3.0,3.5);
\draw[fill=gray] (3.025,3.025) rectangle (3.5,3.5);
\node (a3) at (3.25, 3.25) {1};
\draw[fill=gray] (3.025,0.0) rectangle (3.5,1.5);
\draw[line width=2.0pt] (3.0,3.0) -- (3.5,3.0);
\draw[line width=1.0pt] (3.0,1.5) -- (3.5,1.5);
\draw[line width=2.0pt] (3.0,0.0) -- (3.5,0.0);
\node (a4) at (3.25, 0.75) {1};
\node (b4) at (3.25, 2.25) {3};

\end{tikzpicture}
\caption{Schematic of the communication required before a processor can
  update spins in sector 1.  
  This is a zoomed-in view of the bottom left portion of
  Fig.~\ref{fig:sectoring}.  The current
  values of all spins within the dashed square must
  be known before the processor updates spins in sector 1 (shaded).
  Most of these values are already known, because the spins are
  owned by the processor.  
  But ghost spins in the blue regions are owned by neighboring processors,
  Those other processors must the send the current values to this processor.  
  Similarly, this processor must send the current values of its
  spins in the red regions to neighboring processors.  Once these communication
  operations have been completed, all the processors can concurrently
  advance their spins in sector 1.
} \label{fig:communication}
\end{figure}

Fig.~\ref{fig:communication} is a schematic of the necessary
communication for the same 2d system.  Each processor sends the 
current values of spins associated with the ghost atoms that 
border sector 1.  The thickness of the border region is
equal to the spin-spin interaction cutoff distance.

Note that Fig.~\ref{fig:communication} shows the minimal communication
requirements.
For simplicity and convenience, our current implementation 
makes use of standard communication functions 
in LAMMPS.  All the ghost spins adjacent to the subdomain are updated, 
not only those adjacent to sector 1. 
This simplifies the implementation at the expense of a certain amount of 
unnecessary updating of spins that have not changed value. 
However, in either case, the cost of computation scales as $O((N/P)^{2/3})$, 
with $N$ the number of atoms and $P$ the number of MPI processes, 
differing only in the prefactor\cite{plimpton1995fast}.  In applications where
the communication cost is large relative to computation, extra overall performance
could be achieved by limiting the interprocessor communication to only
those sites adjacent to sector 1.

This pattern of communicating ghost spins then updating a sector is
repeated four times, once for each sector (8 times in 3 dimensions).  
The entire process is then repeated in reverse sector order, at which 
point all of the spins have been updated by a half timestep, according to
Eq. \ref{spin_ST_sectoring}.
Overall, spins in each sector are sequentially updated four times per timestep. 
To accurately integrate the spin dynamics, it is important that
a fixed ordering of spins be used for all four updates. We achieve this
by assigning atoms to sectors once at the start of the
timestep based on their current positions.



 
\subsection{Accuracy of the parallel algorithm}\label{subsec2:accuracy}

\begin{figure}[ht]
\centering
\includegraphics[width=0.95\columnwidth]{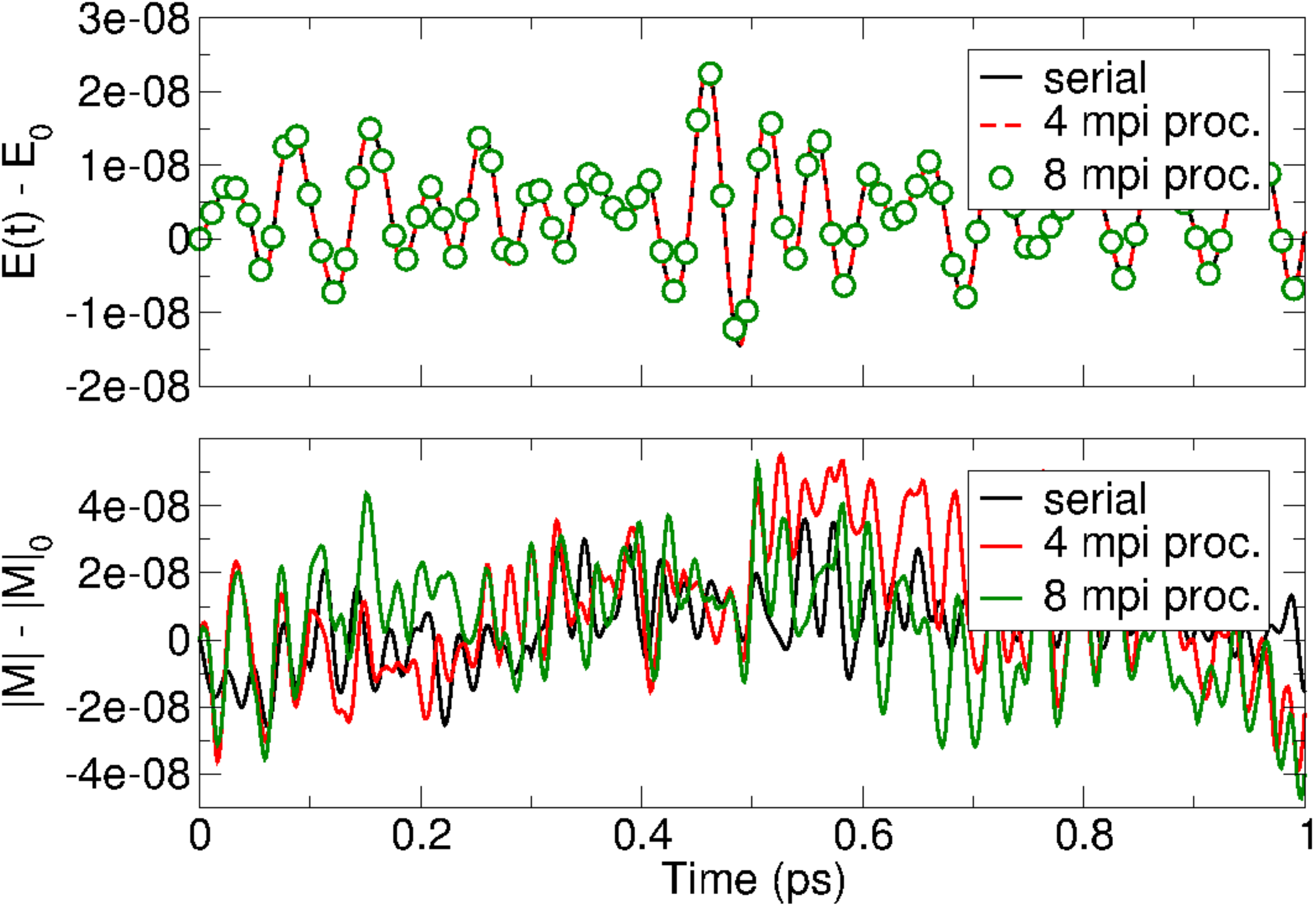}
\caption{NVE SD--MD simulation of a block of 2048 fcc cobalt atoms
  using the sectoring parallel algorithm.  The upper graph plots the
  energy E(t) minus the initial energy E$_0$ as a function of
  simulation time.  The lower graph plots the norm of the total
  magnetization $|{\rm \bm{M}}|$ minus the initial value $|{\rm
    \bm{M}}|_0$.  The black curves are for the serial algorithm, and
  the red and green curves (or dots for the upper graph) are for the
  parallel algorithm.} \label{fig:accuracy_serial_vs_mpi}
\end{figure}

Note that by construction, the parallel algorithm faithfully
reproduces the ST decomposition rule that each spin is updated using
current information for all its neighbor spins.  It differs from the
serial algorithm only in the order in which the global set of spins
are updated.  This order will also differ depending on the number of
processors used to run a simulation.  Two simulations with different
ordering will not evolve identically in a numerical sense, but they
should be produce identical in a statistical sense, since both are
``correct''.  We tested this using the same fcc cobalt model used for
testing the accuracy of the serial algorithm
(Fig.~\ref{fig:timestep_accuracy}).  To allow for multiple processors,
the size of the periodic simulation cell was increased from
17.7~\AA~(500 atoms) to 28.32~\AA~(2048 atoms).  For a $2 \times 2
\times 2$ grid of 8 processors, the width of each sector is 7.08 \AA,
which is larger than the cutoff distance of 6.5~\AA~used by the EAM
potential, and the one of 4.0~\AA~of the magnetic interactions.  
Starting with the same configuration initialized at 300~K,
we ran serial and parallel NVE SD--MD simulations for 1~ps using a
$10^{-4}$ps timestep.  Fig.~\ref{fig:accuracy_serial_vs_mpi} plots the
conservation of total energy and magnetization for the serial
algorithm as well as the parallel algorithm running on 4 and 8 MPI
processes.

The upper graph of Fig.~\ref{fig:accuracy_serial_vs_mpi} shows the
serial and parallel algorithms generate trajectories with tiny energy
fluctuations that are indistinguishable from each other (the red and
black curves overlay), especially relative to the size of the
step-to-step variations in energy due to the numerical integration.
The lower graph of Fig.~\ref{fig:accuracy_serial_vs_mpi} shows there
there are tiny differences between the simulations in the total
magnetization value at a particular timestep.  This is expected,
because the spins are updated in a different order in each simulation,
leading to trajectories that diverge in a numerical sense.  However
the overall step-to-step variation in the magnetization is the same
for all three simulations, indicating the parallel algorithm is
generating a spin trajectory that is statistically equivalent to that
of the serial simulation.

\subsection{Scaling results}\label{subsec3:parallel}

We now evaluate the scaling efficiency of the parallel spin--lattice
algorithm as implemented in LAMMPS, for both strong and weak scaling.
All simulations were performed on a cluster at Sandia consisting of
dual-socket Intel Xeon E5-2695 (Broadwell) CPUs with 36 cores per
node, and an Intel Omnipath interconnect.  To ensure each processor
owns the same number of spins, only 32 cores per node were used for
each of the calculations, though this is not required for general
runs.

Both strong and weak scaling tests were performed.  In both cases, the
parallel efficiency was evaluated by computing a normalized simulation
rate (SR), defined as the number of atoms - steps per CPU second and
per node:
\begin{equation}
  {\rm SR} = \frac{{\rm Steps} \times {\rm Atoms} }
  {{\rm Nodes} \times {\rm Seconds} } 
  \label{pareff}
\end{equation}
and plotted as a function of the number of nodes.

Strong scaling was tested by increasing the number of nodes for a
fixed number of atoms.  Ideally, the computation time should be cut in
half each time the number of nodes is doubled, so that SR should
remain constant.



\begin{figure}[ht]
\centering
\includegraphics[width=0.95\columnwidth]{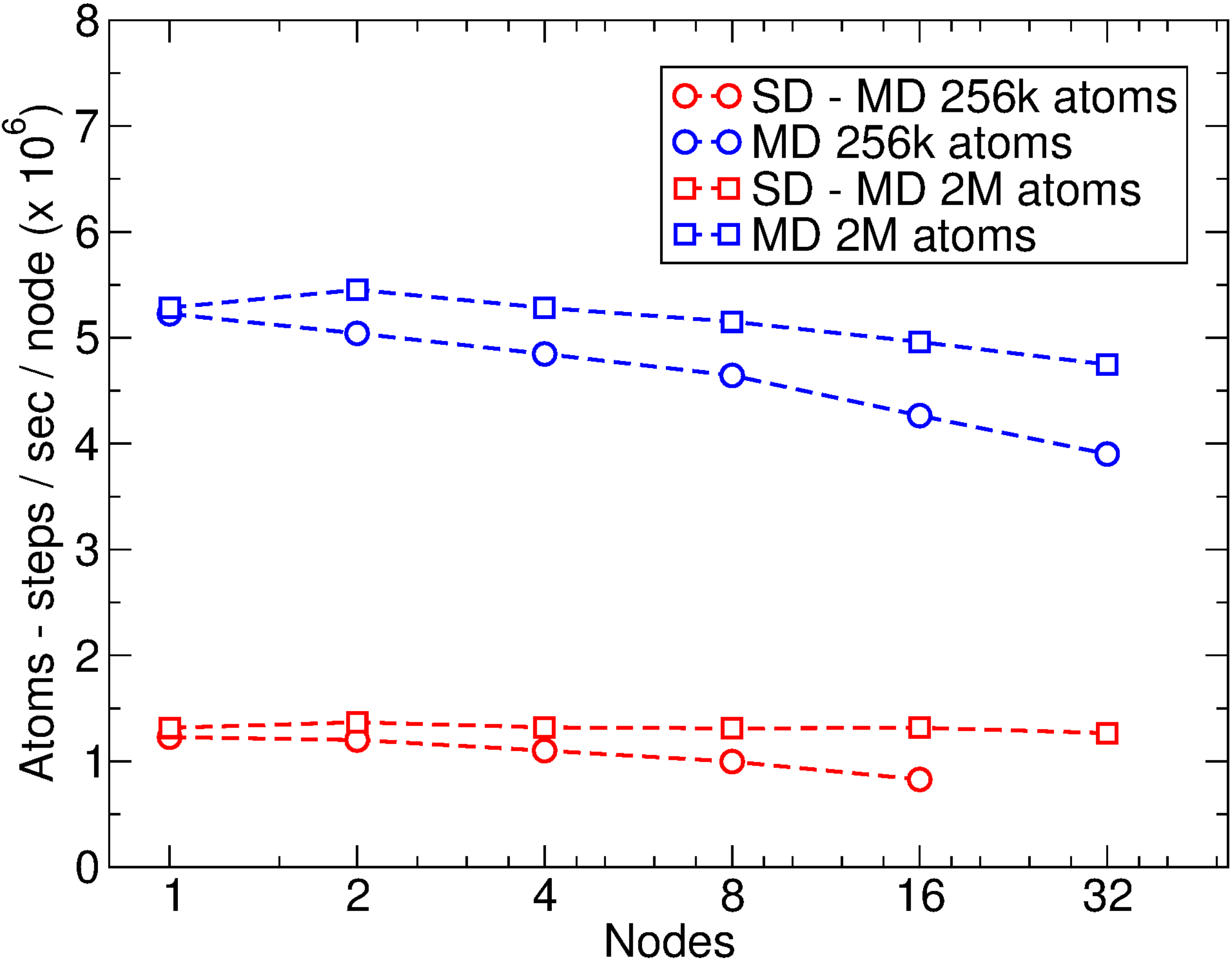}
\caption{Strong-scaling performance of the parallel SD--MD algorithm
   (red) compared to standard MD (blue) for two fixed size problems.
   The smaller system (circles) has 256,000 atoms, the larger
   (squares) has 2,048,000 atoms.  Perfect scaling would be
   horizontal lines.}
\label{fig:strongscaling}
\end{figure}

Each NVE SD--MD simulation combined a mechanical EAM potential with
two magnetic interactions, the exchange energy and a N\'eel anisotropy
(see \ref{app:exchange} and \ref{subapp1:soc} for more details). The
magnetic potentials were parametrized according to
Ref.~\cite{beaujouan_anisotropic_2012}; the EAM potential is described
in Ref.~\cite{pun_embedded-atom_2012}.  Two different problem sizes
were run, the smaller with 256,000 (256K) atoms (a cubic box of $40^3$
fcc unit cells, 4 atoms/cell) and the larger with 2,048,000 (2M) atoms
(2x larger in each dimension).  The SR defined in Eq.~\ref{pareff} was
averaged over a 50 timestep run.  To assess the cost and efficiency of
the SD--MD algorithm, the same simulations were run with only the EAM
potential (no spin variables and without the sectoring algorithm).
Fig.~\ref{fig:strongscaling} shows the results.

Fig.~\ref{fig:strongscaling} indicates the overall computational cost of including
magnetic spin interactions, including the 8x increase in
inter-processor communication to perform the sectoring algorithm, is
about 4-5x that of a standard MD simulation with EAM
potentials. 

For perfect strong scaling, the SR should remain constant as the
number of nodes increases.  As the smaller system is run on more
nodes, the relative cost of interprocessor communication grows as the
atoms/node ratio shrinks, and the SR decreases slowly.  The effect is
less pronounced for the larger problem.  Note that for the smaller
problem, the maximum node count was 16 nodes ($32\times 16=512$ MPI
tasks) to satisfy the requirement that the sector width 
be at least equal to the spin-spin interaction cutoff distance 
(4~\AA).  Also note that the higher computational cost
of the SD--MD model results in somewhat better strong scaling
efficiencies relative to standard MD.

Weak scaling was tested by increasing the size of the system as the
number of nodes was increased, so that the number of atoms per node
(and thus the size of each processor sub-domain) was held fixed.

The same magnetic and EAM interaction models were used as for the
strong scaling runs.  Two different sized systems with 32,000 (32K)
(corresponding to 1,000 atoms per processor), and 256,000 (256K) atoms/node 
(corresponding to 8,000 atoms per processor) were simulated.

\begin{figure}[ht]
\centering
\includegraphics[width=0.95\columnwidth]{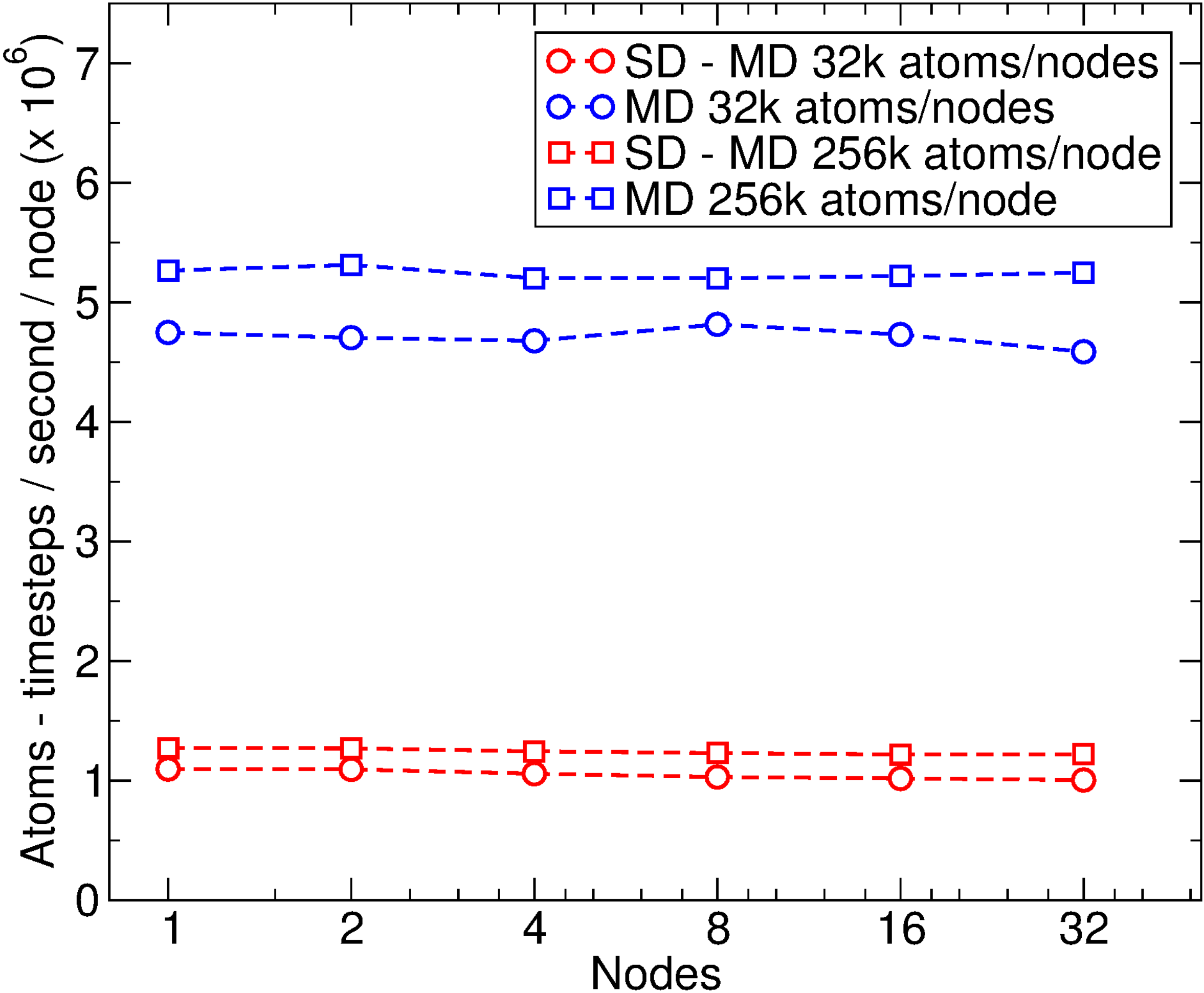}
\caption{Performance of the parallel SD--MD algorithm compared to 
the standard MD algorithm for scaled size
  problems (weak scaling).
  In both cases, the interaction models are the same as that used in
  Fig.~\ref{fig:strongscaling}. 
  The simulation rate (SR in Eq.~\ref{pareff}) is plotted
  as a function of the number of nodes. 
  SD--MD and MD results are colored red and blue, respectively.
  The smaller scaled systems (circles) consisted of 32,000 atoms/node 
  and the larger scaled systems (squares) consisted of 256,000 atoms/node.
} \label{fig:weakscaling}
\end{figure}

Fig.~\ref{fig:weakscaling} shows the results for a number of nodes
ranging from 1 to 64 (corresponding to 2,048 processes).

For ideal weak scaling, the value of SR should
remain constant as the number of nodes increases,
because the amount of work per process remains constant. 
Fig.~\ref{fig:weakscaling} shows that SR indeed decreases 
very little, confirming the efficiency of our SD--MD algorithm. 
Also, we observe that our SD--MD algorithm 
is again approximately 4 to 5 times slower than the standard NVE MD 
calculation with LAMMPS, with the same EAM mechanical potential.

\section{Conclusion}\label{sec:conclusion}

A parallel implementation of coupled magnetic and molecular dynamics
was presented.  This implementation of SD--MD is available as a new
package in LAMMPS.  It allows both coupled and uncoupled simulations
to be performed, i.e. magnetic spins on a fixed lattice of atoms or on
mobile atoms.  Both the serial and parallel versions implement
statistically equivalent symplectic Suzuki-Trotter decompositions of
the spin propagation operators, differing only in the order they
choose to update spins.

The accuracy of the serial version of the coupled algorithm was
analyzed in both the microcanonical (NVE) and canonical (NVT)
statistical ensembles. In the NVE case, care was taken to verify that
both the total internal energy and the norm of the magnetization were
conserved up to second order in the timestep size.  For NVT
simulations testing coupled spin--lattice relaxation, Langevin
thermostats were applied to either the spin subsystem or the lattice
subsystem.

A parallel implementation of the integration algorithm based on the
sectoring or synchronous sublattice method was described in detail.
The statistical equivalence of the parallel algorithm to the serial
algorithm was verified by monitoring the conservation of total energy
and the norm of the magnetization as the number of parallel processors
varied.  The parallel performance of the LAMMPS implementation was
assessed, both for fixed size problems (strong scaling) and scaled
size problems (weak scaling).  In both cases, good performance was
observed for all cases with more than 500 atoms per processor.
Our new SD--MD algorithm with coupled magnetic spin and atomic
lattice dynamics was shown to be only 4 to 5 times 
slower than an analogous NVE MD LAMMPS run 
treating only the motion of the atomic lattice.

This sectoring method has the advantage of being very general, working
for both perfectly ordered particle configurations and disordered
systems, so long as the width of each processor domain is at least
twice the cutoff distance for the short-range spin-spin interactions.

Because the new methods were implemented within the open-source LAMMPS
MD code, they are now available to the scientific community, enabling
a wide variety of coupled spin--lattice simulations to be easily
performed and tested in detail, as well as new models and algorithms
to be implemented.  Applications of this method to the simulation of
magnetoelastic effects in magnetic alloys will be presented in
subsequent publications.

\section*{Acknowledgement}
Sandia National Laboratories is a multimission laboratory managed 
and operated by National Technology and Engineering Solutions of 
Sandia LLC, a wholly owned subsidiary of Honeywell International Inc. 
for the U.S. Department of Energy’s National Nuclear Security 
Administration under contract DE-NA0003525.
J.T. acknowledges financial support through a joint CEA--NNSA
fellowship.  J.T. would also like to thank L. Berger-Vergiat and
L. Bertagna for fruitful conversations, and S. Moore for interesting 
comments and his reviews of this work.





\bibliographystyle{model1-num-names}
\bibliography{MDSD_redaction}

\begin{thebibliography}{87}
\expandafter\ifx\csname natexlab\endcsname\relax\def\natexlab#1{#1}\fi
\providecommand{\bibinfo}[2]{#2}
\ifx\xfnm\relax \def\xfnm[#1]{\unskip,\space#1}\fi
\bibitem[{Smith(2005)}]{smith_smart_2005}
\bibinfo{author}{R.~C. Smith}, \bibinfo{title}{Smart {Material} {Systems},
  {Model} {Development}}, Frontiers in {Applied} {Mathematics},
  \bibinfo{publisher}{SIAM}, \bibinfo{year}{2005}.
\bibitem[{Rossi et~al.(2005)Rossi, Heinonen, and
  MacDonald}]{rossi_dynamics_2005}
\bibinfo{author}{E.~Rossi}, \bibinfo{author}{O.~G. Heinonen},
  \bibinfo{author}{A.~H. MacDonald},
\newblock \bibinfo{title}{Dynamics of magnetization coupled to a thermal bath
  of elastic modes},
\newblock \bibinfo{journal}{Physical Review B} \bibinfo{volume}{72}
  (\bibinfo{year}{2005}).
\bibitem[{Beaurepaire et~al.(1996)Beaurepaire, Merle, Daunois, and
  Bigot}]{beaurepaire_ultrafast_1996}
\bibinfo{author}{E.~Beaurepaire}, \bibinfo{author}{J.-C. Merle},
  \bibinfo{author}{A.~Daunois}, \bibinfo{author}{J.-Y. Bigot},
\newblock \bibinfo{title}{Ultrafast spin dynamics in ferromagnetic nickel},
\newblock \bibinfo{journal}{Physical Review Letters} \bibinfo{volume}{76}
  (\bibinfo{year}{1996}) \bibinfo{pages}{4250}.
\bibitem[{Qian and Vignale(2002)}]{qian_dynamical_2002}
\bibinfo{author}{Z.~Qian}, \bibinfo{author}{G.~Vignale},
\newblock \bibinfo{title}{Dynamical exchange-correlation potentials for an
  electron liquid},
\newblock \bibinfo{journal}{Physical Review B} \bibinfo{volume}{65}
  (\bibinfo{year}{2002}) \bibinfo{pages}{235121}.
\bibitem[{Restrepo and Windl(2012)}]{restrepo_full_2012}
\bibinfo{author}{O.~D. Restrepo}, \bibinfo{author}{W.~Windl},
\newblock \bibinfo{title}{Full {First}-{Principles} {Theory} of {Spin}
  {Relaxation} in {Group}-{IV} {Materials}},
\newblock \bibinfo{journal}{Physical Review Letters} \bibinfo{volume}{109}
  (\bibinfo{year}{2012}).
\bibitem[{Ozdemir~Kart and Cagın(2010)}]{ozdemir_kart_elastic_2010}
\bibinfo{author}{S.~Ozdemir~Kart}, \bibinfo{author}{T.~Cagın},
\newblock \bibinfo{title}{Elastic properties of {Ni}2mnga from first-principles
  calculations},
\newblock \bibinfo{journal}{Journal of Alloys and Compounds}
  \bibinfo{volume}{508} (\bibinfo{year}{2010}) \bibinfo{pages}{177--183}.
\bibitem[{McQuarrie(1976)}]{mcquarrie_statistical_1976}
\bibinfo{author}{D.~A. McQuarrie}, \bibinfo{title}{Statistical {Mechanics}},
  \bibinfo{publisher}{HarperCollins Publishers}, \bibinfo{address}{New York},
  \bibinfo{year}{1976}.
\bibitem[{Aharoni(1996)}]{aharoni_introduction_1996}
\bibinfo{author}{A.~Aharoni}, \bibinfo{title}{Introduction to the {Theory} of
  {Ferromagnetism}}, number~\bibinfo{number}{93} in
  \bibinfo{series}{International series of monographs on physics},
  \bibinfo{publisher}{Oxford Science Publications}, \bibinfo{year}{1996}.
\bibitem[{Antropov et~al.(1995)Antropov, Katsnelson, Van~Schilfgaarde, and
  Harmon}]{antropov_ab_1995}
\bibinfo{author}{V.~P. Antropov}, \bibinfo{author}{M.~I. Katsnelson},
  \bibinfo{author}{M.~Van~Schilfgaarde}, \bibinfo{author}{B.~N. Harmon},
\newblock \bibinfo{title}{Ab {Initio} {Spin} {Dynamics} in {Magnets}},
\newblock \bibinfo{journal}{Physical Review Letters} \bibinfo{volume}{75}
  (\bibinfo{year}{1995}) \bibinfo{pages}{729--732}.
\bibitem[{Antropov et~al.(1996)Antropov, Katsnelson, Harmon, Van~Schilfgaarde,
  and Kusnezov}]{antropov_spin_1996}
\bibinfo{author}{V.~P. Antropov}, \bibinfo{author}{M.~I. Katsnelson},
  \bibinfo{author}{B.~N. Harmon}, \bibinfo{author}{M.~Van~Schilfgaarde},
  \bibinfo{author}{D.~Kusnezov},
\newblock \bibinfo{title}{Spin dynamics in magnets: {Equation} of motion and
  finite temperature effects},
\newblock \bibinfo{journal}{Physical Review B} \bibinfo{volume}{54}
  (\bibinfo{year}{1996}) \bibinfo{pages}{1019}.
\bibitem[{Skubic et~al.(2008)Skubic, Hellsvik, Nordström, and
  Eriksson}]{skubic_method_2008}
\bibinfo{author}{B.~Skubic}, \bibinfo{author}{J.~Hellsvik},
  \bibinfo{author}{L.~Nordström}, \bibinfo{author}{O.~Eriksson},
\newblock \bibinfo{title}{A method for atomistic spin dynamics simulations:
  implementation and examples},
\newblock \bibinfo{journal}{Journal of Physics: Condensed Matter}
  \bibinfo{volume}{20} (\bibinfo{year}{2008}) \bibinfo{pages}{315203}.
\bibitem[{Eriksson et~al.(2017)Eriksson, Bergman, Hellsvik, and
  Bergqvist}]{eriksson_atomistic_2017}
\bibinfo{author}{O.~Eriksson}, \bibinfo{author}{A.~Bergman},
  \bibinfo{author}{J.~Hellsvik}, \bibinfo{author}{L.~Bergqvist},
  \bibinfo{title}{Atomistic spin dynamics: {Foundations} and applications},
  \bibinfo{publisher}{Oxford University Press}, \bibinfo{year}{2017}.
\bibitem[{Radu et~al.(2011)Radu, Vahaplar, Stamm, Kachel, Pontius, Dürr,
  Ostler, Barker, Evans, Chantrell, and {others}}]{radu_transient_2011}
\bibinfo{author}{I.~Radu}, \bibinfo{author}{K.~Vahaplar},
  \bibinfo{author}{C.~Stamm}, \bibinfo{author}{T.~Kachel},
  \bibinfo{author}{N.~Pontius}, \bibinfo{author}{H.~Dürr},
  \bibinfo{author}{T.~Ostler}, \bibinfo{author}{J.~Barker},
  \bibinfo{author}{R.~Evans}, \bibinfo{author}{R.~Chantrell},
  \bibinfo{author}{{others}},
\newblock \bibinfo{title}{Transient ferromagnetic-like state mediating
  ultrafast reversal of antiferromagnetically coupled spins},
\newblock \bibinfo{journal}{Nature} \bibinfo{volume}{472}
  (\bibinfo{year}{2011}) \bibinfo{pages}{205}.
\bibitem[{Dupé et~al.(2014)Dupé, Hoffmann, Paillard, and
  Heinze}]{dupe_tailoring_2014}
\bibinfo{author}{B.~Dupé}, \bibinfo{author}{M.~Hoffmann},
  \bibinfo{author}{C.~Paillard}, \bibinfo{author}{S.~Heinze},
\newblock \bibinfo{title}{Tailoring magnetic skyrmions in ultra-thin transition
  metal films},
\newblock \bibinfo{journal}{Nature communications} \bibinfo{volume}{5}
  (\bibinfo{year}{2014}) \bibinfo{pages}{4030}.
\bibitem[{Tranchida et~al.(2016)Tranchida, Thibaudeau, and
  Nicolis}]{tranchida_functional_2016}
\bibinfo{author}{J.~Tranchida}, \bibinfo{author}{P.~Thibaudeau},
  \bibinfo{author}{S.~Nicolis},
\newblock \bibinfo{title}{A functional calculus for the magnetization
  dynamics},
\newblock \bibinfo{journal}{arXiv preprint arXiv:1606.02137}
  (\bibinfo{year}{2016}).
\bibitem[{Ma et~al.(2008)Ma, Woo, and Dudarev}]{ma_large-scale_2008}
\bibinfo{author}{P.-W. Ma}, \bibinfo{author}{C.~H. Woo}, \bibinfo{author}{S.~L.
  Dudarev},
\newblock \bibinfo{title}{Large-scale simulation of the spin-lattice dynamics
  in ferromagnetic iron},
\newblock \bibinfo{journal}{Physical Review B} \bibinfo{volume}{78}
  (\bibinfo{year}{2008}) \bibinfo{pages}{024434}.
\bibitem[{Ma et~al.(2016)Ma, Dudarev, and Woo}]{ma_spilady:_2016}
\bibinfo{author}{P.-W. Ma}, \bibinfo{author}{S.~L. Dudarev},
  \bibinfo{author}{C.~H. Woo},
\newblock \bibinfo{title}{{SPILADY}: {A} parallel {CPU} and {GPU} code for
  spin–lattice magnetic molecular dynamics simulations},
\newblock \bibinfo{journal}{Computer Physics Communications}
  \bibinfo{volume}{207} (\bibinfo{year}{2016}) \bibinfo{pages}{350--361}.
\bibitem[{Beaujouan et~al.(2012)Beaujouan, Thibaudeau, and
  Barreteau}]{beaujouan_anisotropic_2012}
\bibinfo{author}{D.~Beaujouan}, \bibinfo{author}{P.~Thibaudeau},
  \bibinfo{author}{C.~Barreteau},
\newblock \bibinfo{title}{Anisotropic magnetic molecular dynamics of cobalt
  nanowires},
\newblock \bibinfo{journal}{Physical Review B} \bibinfo{volume}{86}
  (\bibinfo{year}{2012}) \bibinfo{pages}{174409}.
\bibitem[{Perera et~al.(2016)Perera, Eisenbach, Nicholson, Stocks, and
  Landau}]{perera_reinventing_2016}
\bibinfo{author}{D.~Perera}, \bibinfo{author}{M.~Eisenbach},
  \bibinfo{author}{D.~M. Nicholson}, \bibinfo{author}{G.~M. Stocks},
  \bibinfo{author}{D.~P. Landau},
\newblock \bibinfo{title}{Reinventing atomistic magnetic simulations with
  spin-orbit coupling},
\newblock \bibinfo{journal}{Physical Review B} \bibinfo{volume}{93}
  (\bibinfo{year}{2016}) \bibinfo{pages}{060402}.
\bibitem[{Perera et~al.(2017)Perera, Nicholson, Eisenbach, Stocks, and
  Landau}]{perera_collective_2017}
\bibinfo{author}{D.~Perera}, \bibinfo{author}{D.~M. Nicholson},
  \bibinfo{author}{M.~Eisenbach}, \bibinfo{author}{G.~M. Stocks},
  \bibinfo{author}{D.~P. Landau},
\newblock \bibinfo{title}{Collective dynamics in atomistic models with coupled
  translational and spin degrees of freedom},
\newblock \bibinfo{journal}{Physical Review B} \bibinfo{volume}{95}
  (\bibinfo{year}{2017}) \bibinfo{pages}{014431}.
\bibitem[{Plimpton et~al.(2007)Plimpton, Crozier, and
  Thompson}]{plimpton_lammps-large-scale_2007}
\bibinfo{author}{S.~Plimpton}, \bibinfo{author}{P.~Crozier},
  \bibinfo{author}{A.~Thompson},
\newblock \bibinfo{title}{{LAMMPS}-{Large}-scale {Atomic}/{Molecular}
  {Massively} {Parallel} {Simulator}},
\newblock \bibinfo{journal}{Sandia National Laboratories} \bibinfo{volume}{18}
  (\bibinfo{year}{2007}).
\bibitem[{Blanes et~al.(2009)Blanes, Casas, Oteo, and Ros}]{blanes_magnus_2009}
\bibinfo{author}{S.~Blanes}, \bibinfo{author}{F.~Casas}, \bibinfo{author}{J.~A.
  Oteo}, \bibinfo{author}{J.~Ros},
\newblock \bibinfo{title}{The {Magnus} expansion and some of its applications},
\newblock \bibinfo{journal}{Physics Reports} \bibinfo{volume}{470}
  (\bibinfo{year}{2009}) \bibinfo{pages}{151--238}.
\bibitem[{Shim and Amar(2005)}]{shim_semirigorous_2005}
\bibinfo{author}{Y.~Shim}, \bibinfo{author}{J.~G. Amar},
\newblock \bibinfo{title}{Semirigorous synchronous sublattice algorithm for
  parallel kinetic {Monte} {Carlo} simulations of thin film growth},
\newblock \bibinfo{journal}{Physical Review B} \bibinfo{volume}{71}
  (\bibinfo{year}{2005}) \bibinfo{pages}{125432}.
\bibitem[{Bacry(1962)}]{bacry_thomass_1962}
\bibinfo{author}{H.~Bacry},
\newblock \bibinfo{title}{Thomas's classical theory of spin},
\newblock \bibinfo{journal}{Il Nuovo Cimento} \bibinfo{volume}{26}
  (\bibinfo{year}{1962}) \bibinfo{pages}{1164--1172}.
\bibitem[{Thomas(1926)}]{thomas_motion_1926}
\bibinfo{author}{L.~H. Thomas},
\newblock \bibinfo{title}{The {Motion} of the {Spinning} {Electron}},
\newblock \bibinfo{journal}{Nature} \bibinfo{volume}{117}
  (\bibinfo{year}{1926}) \bibinfo{pages}{514--514}.
\bibitem[{Thomas(1927)}]{thomas_i._1927}
\bibinfo{author}{L.~Thomas},
\newblock \bibinfo{title}{I. {The} kinematics of an electron with an axis},
\newblock \bibinfo{journal}{The London, Edinburgh, and Dublin Philosophical
  Magazine and Journal of Science} \bibinfo{volume}{3} (\bibinfo{year}{1927})
  \bibinfo{pages}{1--22}.
\bibitem[{Weiss(2012)}]{weiss_quantum_2012}
\bibinfo{author}{U.~Weiss}, \bibinfo{title}{Quantum dissipative systems},
  \bibinfo{publisher}{World Scientific}, \bibinfo{address}{New Jersey, NJ},
  \bibinfo{edition}{4. ed} edition, \bibinfo{year}{2012}.
\bibitem[{Wieser(2015)}]{wieser_description_2015}
\bibinfo{author}{R.~Wieser},
\newblock \bibinfo{title}{Description of a dissipative quantum spin dynamics
  with a {Landau}-{Lifshitz}/{Gilbert} like damping and complete derivation of
  the classical {Landau}-{Lifshitz} equation},
\newblock \bibinfo{journal}{The European Physical Journal B}
  \bibinfo{volume}{88} (\bibinfo{year}{2015}).
\bibitem[{Gilbert(2004)}]{gilbert_phenomenological_2004}
\bibinfo{author}{T.~L. Gilbert},
\newblock \bibinfo{title}{A phenomenological theory of damping in ferromagnetic
  materials},
\newblock \bibinfo{journal}{IEEE Transactions on Magnetics}
  \bibinfo{volume}{40} (\bibinfo{year}{2004}) \bibinfo{pages}{3443--3449}.
\bibitem[{Ebert et~al.(2011)Ebert, Mankovsky, Ködderitzsch, and
  Kelly}]{ebert_ab_2011}
\bibinfo{author}{H.~Ebert}, \bibinfo{author}{S.~Mankovsky},
  \bibinfo{author}{D.~Ködderitzsch}, \bibinfo{author}{P.~J. Kelly},
\newblock \bibinfo{title}{\textit{{Ab}} {Initio} {Calculation} of the {Gilbert}
  {Damping} {Parameter} via the {Linear} {Response} {Formalism}},
\newblock \bibinfo{journal}{Physical Review Letters} \bibinfo{volume}{107}
  (\bibinfo{year}{2011}).
\bibitem[{Thibaudeau and Beaujouan(2012)}]{thibaudeau_thermostatting_2012}
\bibinfo{author}{P.~Thibaudeau}, \bibinfo{author}{D.~Beaujouan},
\newblock \bibinfo{title}{Thermostatting the atomic spin dynamics from
  controlled demons},
\newblock \bibinfo{journal}{Physica A: Statistical Mechanics and its
  Applications} \bibinfo{volume}{391} (\bibinfo{year}{2012})
  \bibinfo{pages}{1963--1971}.
\bibitem[{Kittel and Fong(1987)}]{kittel_quantum_1987}
\bibinfo{author}{C.~Kittel}, \bibinfo{author}{C.-y. Fong},
  \bibinfo{title}{Quantum theory of solids: includes solutions appendix,
  prepared by {C}.{Y}. {Fong}}, \bibinfo{publisher}{Wiley},
  \bibinfo{address}{New York}, \bibinfo{edition}{2., rev. print} edition,
  \bibinfo{year}{1987}.
\bibitem[{Victora and MacLaren(1993)}]{victora_theory_1993}
\bibinfo{author}{R.~H. Victora}, \bibinfo{author}{J.~M. MacLaren},
\newblock \bibinfo{title}{Theory of magnetic interface anisotropy},
\newblock \bibinfo{journal}{Physical Review B} \bibinfo{volume}{47}
  (\bibinfo{year}{1993}) \bibinfo{pages}{11583--11586}.
\bibitem[{Néel(1954)}]{neel_approche_1954}
\bibinfo{author}{L.~Néel},
\newblock \bibinfo{title}{L'approche à la saturation de la magnétostriction},
\newblock \bibinfo{journal}{Journal de Physique et le Radium}
  \bibinfo{volume}{15} (\bibinfo{year}{1954}) \bibinfo{pages}{376--378}.
\bibitem[{Bruno(1988)}]{bruno_magnetic_1988}
\bibinfo{author}{P.~Bruno},
\newblock \bibinfo{title}{Magnetic surface anisotropy of cobalt and surface
  roughness effects within {Neel}'s model},
\newblock \bibinfo{journal}{Journal of Physics F: Metal Physics}
  \bibinfo{volume}{18} (\bibinfo{year}{1988}) \bibinfo{pages}{1291--1298}.
\bibitem[{Dudarev and Derlet(2005)}]{dudarev_`magneticinteratomic_2005}
\bibinfo{author}{S.~Dudarev}, \bibinfo{author}{P.~Derlet},
\newblock \bibinfo{title}{A `magnetic'interatomic potential for molecular
  dynamics simulations},
\newblock \bibinfo{journal}{Journal of Physics: Condensed Matter}
  \bibinfo{volume}{17} (\bibinfo{year}{2005}) \bibinfo{pages}{7097}.
\bibitem[{Thibaudeau and Gale(2008)}]{thibaudeau_embedded-atom_2008}
\bibinfo{author}{P.~Thibaudeau}, \bibinfo{author}{J.~Gale},
\newblock \bibinfo{title}{An embedded-atom method model for liquid {Co}, {Nb},
  {Zr} and supercooled binary alloys},
\newblock \bibinfo{journal}{arXiv preprint arXiv:0809.0198}
  (\bibinfo{year}{2008}).
\bibitem[{Yang and Hirschfelder(1980)}]{yang_generalizations_1980}
\bibinfo{author}{K.-H. Yang}, \bibinfo{author}{J.~O. Hirschfelder},
\newblock \bibinfo{title}{Generalizations of classical {Poisson} brackets to
  include spin},
\newblock \bibinfo{journal}{Physical Review A} \bibinfo{volume}{22}
  (\bibinfo{year}{1980}) \bibinfo{pages}{1814}.
\bibitem[{Tuckerman(2010)}]{tuckerman_statistical_2010}
\bibinfo{author}{M.~Tuckerman}, \bibinfo{title}{Statistical mechanics: theory
  and molecular simulation}, \bibinfo{publisher}{Oxford University Press},
  \bibinfo{year}{2010}.
\bibitem[{Frenkel and Smit(2002)}]{frenkel_understanding_2002}
\bibinfo{author}{D.~Frenkel}, \bibinfo{author}{B.~Smit},
  \bibinfo{title}{Understanding molecular simulation: from algorithms to
  applications}, \bibinfo{publisher}{Academic Press}, \bibinfo{address}{San
  Diego}, \bibinfo{edition}{2} edition, \bibinfo{year}{2002}.
\bibitem[{Omelyan et~al.(2001)Omelyan, Mryglod, and
  Folk}]{omelyan_algorithm_2001}
\bibinfo{author}{I.~Omelyan}, \bibinfo{author}{I.~Mryglod},
  \bibinfo{author}{R.~Folk},
\newblock \bibinfo{title}{Algorithm for molecular dynamics simulations of spin
  liquids},
\newblock \bibinfo{journal}{Physical Review Letters} \bibinfo{volume}{86}
  (\bibinfo{year}{2001}) \bibinfo{pages}{898}.
\bibitem[{Dullweber et~al.(1997)Dullweber, Leimkuhler, and
  McLachlan}]{dullweber_symplectic_1997}
\bibinfo{author}{A.~Dullweber}, \bibinfo{author}{B.~Leimkuhler},
  \bibinfo{author}{R.~McLachlan},
\newblock \bibinfo{title}{Symplectic splitting methods for rigid body molecular
  dynamics},
\newblock \bibinfo{journal}{The Journal of Chemical Physics}
  \bibinfo{volume}{107} (\bibinfo{year}{1997}) \bibinfo{pages}{5840--5851}.
\bibitem[{Andersen(1980)}]{andersen_molecular_1980}
\bibinfo{author}{H.~C. Andersen},
\newblock \bibinfo{title}{Molecular dynamics simulations at constant pressure
  and/or temperature},
\newblock \bibinfo{journal}{The Journal of Chemical Physics}
  \bibinfo{volume}{72} (\bibinfo{year}{1980}) \bibinfo{pages}{2384--2393}.
\bibitem[{Nosé(1984)}]{nose_unified_1984}
\bibinfo{author}{S.~Nosé},
\newblock \bibinfo{title}{A unified formulation of the constant temperature
  molecular dynamics methods},
\newblock \bibinfo{journal}{The Journal of Chemical Physics}
  \bibinfo{volume}{81} (\bibinfo{year}{1984}) \bibinfo{pages}{511--519}.
\bibitem[{Hoover(1985)}]{hoover_canonical_1985}
\bibinfo{author}{W.~G. Hoover},
\newblock \bibinfo{title}{Canonical dynamics: {Equilibrium} phase-space
  distributions},
\newblock \bibinfo{journal}{Physical Review A} \bibinfo{volume}{31}
  (\bibinfo{year}{1985}) \bibinfo{pages}{1695--1697}.
\bibitem[{Feller et~al.(1995)Feller, Zhang, Pastor, and
  Brooks}]{feller_constant_1995}
\bibinfo{author}{S.~E. Feller}, \bibinfo{author}{Y.~Zhang},
  \bibinfo{author}{R.~W. Pastor}, \bibinfo{author}{B.~R. Brooks},
\newblock \bibinfo{title}{Constant pressure molecular dynamics simulation:
  {The} {Langevin} piston method},
\newblock \bibinfo{journal}{The Journal of Chemical Physics}
  \bibinfo{volume}{103} (\bibinfo{year}{1995}) \bibinfo{pages}{4613--4621}.
\bibitem[{Hoogerbrugge and Koelman(1992)}]{hoogerbrugge_simulating_1992}
\bibinfo{author}{P.~J. Hoogerbrugge}, \bibinfo{author}{J.~M. V.~A. Koelman},
\newblock \bibinfo{title}{Simulating {Microscopic} {Hydrodynamic} {Phenomena}
  with {Dissipative} {Particle} {Dynamics}},
\newblock \bibinfo{journal}{Europhysics Letters} \bibinfo{volume}{19}
  (\bibinfo{year}{1992}) \bibinfo{pages}{155--160}.
\bibitem[{Groot and Warren(1997)}]{groot_dissipative_1997}
\bibinfo{author}{R.~D. Groot}, \bibinfo{author}{P.~B. Warren},
\newblock \bibinfo{title}{Dissipative particle dynamics: {Bridging} the gap
  between atomistic and mesoscopic simulation},
\newblock \bibinfo{journal}{The Journal of Chemical Physics}
  \bibinfo{volume}{107} (\bibinfo{year}{1997}) \bibinfo{pages}{4423--4435}.
\bibitem[{Tuckerman et~al.(2001)Tuckerman, Liu, Ciccotti, and
  Martyna}]{tuckerman_non-hamiltonian_2001}
\bibinfo{author}{M.~E. Tuckerman}, \bibinfo{author}{Y.~Liu},
  \bibinfo{author}{G.~Ciccotti}, \bibinfo{author}{G.~J. Martyna},
\newblock \bibinfo{title}{Non-{Hamiltonian} molecular dynamics: {Generalizing}
  {Hamiltonian} phase space principles to non-{Hamiltonian} systems},
\newblock \bibinfo{journal}{The Journal of Chemical Physics}
  \bibinfo{volume}{115} (\bibinfo{year}{2001}) \bibinfo{pages}{1678--1702}.
\bibitem[{Néel(1949)}]{neel_influence_1949}
\bibinfo{author}{L.~Néel},
\newblock \bibinfo{title}{Influence des fluctuations thermiques sur
  l'aimantation de grains ferromagnétiques très fins},
\newblock \bibinfo{journal}{Compte Rendus de l'Académie des Sciences}
  \bibinfo{volume}{228} (\bibinfo{year}{1949}) \bibinfo{pages}{664}.
\bibitem[{Brown(1963)}]{brown_thermal_1963}
\bibinfo{author}{W.~F. Brown},
\newblock \bibinfo{title}{Thermal {Fluctuations} of a {Single}-{Domain}
  {Particle}},
\newblock \bibinfo{journal}{Physical Review} \bibinfo{volume}{130}
  (\bibinfo{year}{1963}) \bibinfo{pages}{1677--1686}.
\bibitem[{García-Palacios and
  Lázaro(1998)}]{garcia-palacios_langevin-dynamics_1998}
\bibinfo{author}{J.~L. García-Palacios}, \bibinfo{author}{F.~J. Lázaro},
\newblock \bibinfo{title}{Langevin-dynamics study of the dynamical properties
  of small magnetic particles},
\newblock \bibinfo{journal}{Physical Review B} \bibinfo{volume}{58}
  (\bibinfo{year}{1998}) \bibinfo{pages}{14937}.
\bibitem[{Mayergoyz et~al.(2009)Mayergoyz, Bertotti, and
  Serpico}]{mayergoyz_nonlinear_2009}
\bibinfo{author}{I.~D. Mayergoyz}, \bibinfo{author}{G.~Bertotti},
  \bibinfo{author}{C.~Serpico}, \bibinfo{title}{Nonlinear magnetization
  dynamics in nanosystems}, \bibinfo{publisher}{Elsevier},
  \bibinfo{year}{2009}.
\bibitem[{d'Aquino et~al.(2006)d'Aquino, Serpico, Coppola, Mayergoyz, and
  Bertotti}]{daquino_midpoint_2006}
\bibinfo{author}{M.~d'Aquino}, \bibinfo{author}{C.~Serpico},
  \bibinfo{author}{G.~Coppola}, \bibinfo{author}{I.~Mayergoyz},
  \bibinfo{author}{G.~Bertotti},
\newblock \bibinfo{title}{Midpoint numerical technique for stochastic
  {Landau}-{Lifshitz}-{Gilbert} dynamics},
\newblock \bibinfo{journal}{Journal of Applied Physics} \bibinfo{volume}{99}
  (\bibinfo{year}{2006}) \bibinfo{pages}{08B905}.
\bibitem[{Aron et~al.(2014)Aron, Barci, Cugliandolo, Arenas, and
  Lozano}]{aron_magnetization_2014}
\bibinfo{author}{C.~Aron}, \bibinfo{author}{D.~G. Barci},
  \bibinfo{author}{L.~F. Cugliandolo}, \bibinfo{author}{Z.~G. Arenas},
  \bibinfo{author}{G.~S. Lozano},
\newblock \bibinfo{title}{Magnetization dynamics: path-integral formalism for
  the stochastic {Landau}–{Lifshitz}–{Gilbert} equation},
\newblock \bibinfo{journal}{Journal of Statistical Mechanics: Theory and
  Experiment} \bibinfo{volume}{2014} (\bibinfo{year}{2014})
  \bibinfo{pages}{P09008}.
\bibitem[{Rapaport(2004)}]{rapaport_art_2004}
\bibinfo{author}{D.~C. Rapaport}, \bibinfo{title}{The {Art} of {Molecular}
  {Dynamics} {Simulation}}, \bibinfo{publisher}{Cambridge University Press},
  \bibinfo{edition}{2nd} edition, \bibinfo{year}{2004}.
\bibitem[{Rugh(1997)}]{rugh_dynamical_1997}
\bibinfo{author}{H.~H. Rugh},
\newblock \bibinfo{title}{Dynamical approach to temperature},
\newblock \bibinfo{journal}{Physical Review Letters} \bibinfo{volume}{78}
  (\bibinfo{year}{1997}) \bibinfo{pages}{772}.
\bibitem[{Rugh(1998)}]{rugh_geometric_1998}
\bibinfo{author}{H.~H. Rugh},
\newblock \bibinfo{title}{A geometric, dynamical approach to thermodynamics},
\newblock \bibinfo{journal}{Journal of Physics A: Mathematical and General}
  \bibinfo{volume}{31} (\bibinfo{year}{1998}) \bibinfo{pages}{7761}.
\bibitem[{Jepps et~al.(2000)Jepps, Ayton, and Evans}]{jepps_microscopic_2000}
\bibinfo{author}{O.~G. Jepps}, \bibinfo{author}{G.~Ayton},
  \bibinfo{author}{D.~J. Evans},
\newblock \bibinfo{title}{Microscopic expressions for the thermodynamic
  temperature},
\newblock \bibinfo{journal}{Physical Review E} \bibinfo{volume}{62}
  (\bibinfo{year}{2000}) \bibinfo{pages}{4757--4763}.
\bibitem[{Nurdin and Schotte(2000)}]{nurdin_dynamical_2000}
\bibinfo{author}{W.~B. Nurdin}, \bibinfo{author}{K.-D. Schotte},
\newblock \bibinfo{title}{Dynamical temperature for spin systems},
\newblock \bibinfo{journal}{Physical Review E} \bibinfo{volume}{61}
  (\bibinfo{year}{2000}) \bibinfo{pages}{3579}.
\bibitem[{Ma et~al.(2010)Ma, Dudarev, Semenov, and Woo}]{ma_temperature_2010}
\bibinfo{author}{P.-W. Ma}, \bibinfo{author}{S.~L. Dudarev},
  \bibinfo{author}{A.~A. Semenov}, \bibinfo{author}{C.~H. Woo},
\newblock \bibinfo{title}{Temperature for a dynamic spin ensemble},
\newblock \bibinfo{journal}{Physical Review E} \bibinfo{volume}{82}
  (\bibinfo{year}{2010}).
\bibitem[{Ma et~al.(2012)Ma, Dudarev, and Woo}]{ma_spin-lattice-electron_2012}
\bibinfo{author}{P.-W. Ma}, \bibinfo{author}{S.~Dudarev},
  \bibinfo{author}{C.~Woo},
\newblock \bibinfo{title}{Spin-lattice-electron dynamics simulations of
  magnetic materials},
\newblock \bibinfo{journal}{Physical Review B} \bibinfo{volume}{85}
  (\bibinfo{year}{2012}) \bibinfo{pages}{184301}.
\bibitem[{Atxitia et~al.(2009)Atxitia, Chubykalo-Fesenko, Chantrell, Nowak, and
  Rebei}]{atxitia_ultrafast_2009}
\bibinfo{author}{U.~Atxitia}, \bibinfo{author}{O.~Chubykalo-Fesenko},
  \bibinfo{author}{R.~W. Chantrell}, \bibinfo{author}{U.~Nowak},
  \bibinfo{author}{A.~Rebei},
\newblock \bibinfo{title}{Ultrafast {Spin} {Dynamics}: {The} {Effect} of
  {Colored} {Noise}},
\newblock \bibinfo{journal}{Physical Review Letters} \bibinfo{volume}{102}
  (\bibinfo{year}{2009}).
\bibitem[{Bose and Trimper(2010)}]{bose_correlation_2010}
\bibinfo{author}{T.~Bose}, \bibinfo{author}{S.~Trimper},
\newblock \bibinfo{title}{Correlation effects in the stochastic
  {Landau}-{Lifshitz}-{Gilbert} equation},
\newblock \bibinfo{journal}{Physical Review B} \bibinfo{volume}{81}
  (\bibinfo{year}{2010}).
\bibitem[{Tranchida et~al.(2016{\natexlab{a}})Tranchida, Thibaudeau, and
  Nicolis}]{tranchida_closing_2016}
\bibinfo{author}{J.~Tranchida}, \bibinfo{author}{P.~Thibaudeau},
  \bibinfo{author}{S.~Nicolis},
\newblock \bibinfo{title}{Closing the hierarchy for non-{Markovian}
  magnetization dynamics},
\newblock \bibinfo{journal}{Physica B: Condensed Matter} \bibinfo{volume}{486}
  (\bibinfo{year}{2016}{\natexlab{a}}) \bibinfo{pages}{57--59}.
\bibitem[{Tranchida et~al.(2016{\natexlab{b}})Tranchida, Thibaudeau, and
  Nicolis}]{tranchida_colored-noise_2016}
\bibinfo{author}{J.~Tranchida}, \bibinfo{author}{P.~Thibaudeau},
  \bibinfo{author}{S.~Nicolis},
\newblock \bibinfo{title}{Colored-noise magnetization dynamics: from weakly to
  strongly correlated noise},
\newblock \bibinfo{journal}{IEEE Transactions on Magnetics}
  \bibinfo{volume}{52} (\bibinfo{year}{2016}{\natexlab{b}})
  \bibinfo{pages}{1300504}.
\bibitem[{Pun and Mishin(2012)}]{pun_embedded-atom_2012}
\bibinfo{author}{G.~P. Pun}, \bibinfo{author}{Y.~Mishin},
\newblock \bibinfo{title}{Embedded-atom potential for hcp and fcc cobalt},
\newblock \bibinfo{journal}{Physical Review B} \bibinfo{volume}{86}
  (\bibinfo{year}{2012}) \bibinfo{pages}{134116}.
\bibitem[{Krech et~al.(1998)Krech, Bunker, and Landau}]{krech_fast_1998}
\bibinfo{author}{M.~Krech}, \bibinfo{author}{A.~Bunker},
  \bibinfo{author}{D.~Landau},
\newblock \bibinfo{title}{Fast spin dynamics algorithms for classical spin
  systems},
\newblock \bibinfo{journal}{Computer Physics Communications}
  \bibinfo{volume}{111} (\bibinfo{year}{1998}) \bibinfo{pages}{1--13}.
\bibitem[{Ma and Woo(2009)}]{ma_parallel_2009}
\bibinfo{author}{P.-W. Ma}, \bibinfo{author}{C.~Woo},
\newblock \bibinfo{title}{Parallel algorithm for spin and spin-lattice dynamics
  simulations},
\newblock \bibinfo{journal}{Physical Review E} \bibinfo{volume}{79}
  (\bibinfo{year}{2009}) \bibinfo{pages}{046703}.
\bibitem[{Plimpton et~al.(2009)Plimpton, Battaile, Chandross, Holm, Thompson,
  Tikare, Wagner, Webb, Zhou, Cardona, and {others}}]{plimpton_crossing_2009}
\bibinfo{author}{S.~Plimpton}, \bibinfo{author}{C.~Battaile},
  \bibinfo{author}{M.~Chandross}, \bibinfo{author}{L.~Holm},
  \bibinfo{author}{A.~Thompson}, \bibinfo{author}{V.~Tikare},
  \bibinfo{author}{G.~Wagner}, \bibinfo{author}{E.~Webb},
  \bibinfo{author}{X.~Zhou}, \bibinfo{author}{C.~G. Cardona},
  \bibinfo{author}{{others}}, \bibinfo{title}{Crossing the mesoscale no-man's
  land via parallel kinetic {Monte} {Carlo}}, \bibinfo{type}{Technical Report}
  \bibinfo{number}{SAND2009-6226}, Sandia National Laboratories,
  \bibinfo{year}{2009}.
\bibitem[{Plimpton(1995)}]{plimpton1995fast}
\bibinfo{author}{S.~Plimpton},
\newblock \bibinfo{title}{Fast parallel algorithms for short-range molecular
  dynamics},
\newblock \bibinfo{journal}{Journal of computational physics}
  \bibinfo{volume}{117} (\bibinfo{year}{1995}) \bibinfo{pages}{1--19}.
\bibitem[{Van~Leeuwen(1921)}]{van_leeuwen_problemes_1921}
\bibinfo{author}{H.-J. Van~Leeuwen},
\newblock \bibinfo{title}{Problèmes de la théorie électronique du
  magnétisme},
\newblock \bibinfo{journal}{Journal de Physique et le Radium}
  \bibinfo{volume}{2} (\bibinfo{year}{1921}) \bibinfo{pages}{361--377}.
\bibitem[{Duck and Sudarshan(1998)}]{duck_pauli_1998}
\bibinfo{author}{I.~Duck}, \bibinfo{author}{E.~C.~G. Sudarshan},
  \bibinfo{title}{Pauli and the {Spin}-{Statistics} {Theorem}},
  \bibinfo{publisher}{World Scientific}, \bibinfo{year}{1998}.
\bibitem[{Kaneyoshi(1992)}]{kaneyoshi_introduction_1992}
\bibinfo{author}{T.~Kaneyoshi}, \bibinfo{title}{Introduction to amorphous
  magnets}, \bibinfo{publisher}{World Scientific},
  \bibinfo{address}{Singapore}, \bibinfo{year}{1992}.
\bibitem[{Yosida(1996)}]{yosida_theory_1996}
\bibinfo{author}{K.~Yosida}, \bibinfo{title}{Theory of magnetism}, number
  \bibinfo{number}{122} in \bibinfo{series}{Springer series in solid-state
  sciences}, \bibinfo{publisher}{Springer}, \bibinfo{address}{Berlin ; New
  York}, \bibinfo{year}{1996}.
\bibitem[{Pajda et~al.(2001)Pajda, Kudrnovsky, Turek, Drchal, and
  Bruno}]{pajda_ab_2001}
\bibinfo{author}{M.~Pajda}, \bibinfo{author}{J.~Kudrnovsky},
  \bibinfo{author}{I.~Turek}, \bibinfo{author}{V.~Drchal},
  \bibinfo{author}{P.~Bruno},
\newblock \bibinfo{title}{Ab initio calculations of exchange interactions,
  spin-wave stiffness constants, and {Curie} temperatures of {Fe}, {Co}, and
  {Ni}},
\newblock \bibinfo{journal}{Physical Review B} \bibinfo{volume}{64}
  (\bibinfo{year}{2001}) \bibinfo{pages}{174402}.
\bibitem[{Jeong et~al.(2012)Jeong, Goremychkin, Guidi, Nakajima, Jeon, Kim,
  Furukawa, Kim, Lee, Kiryukhin, and {others}}]{jeong_spin_2012}
\bibinfo{author}{J.~Jeong}, \bibinfo{author}{E.~Goremychkin},
  \bibinfo{author}{T.~Guidi}, \bibinfo{author}{K.~Nakajima},
  \bibinfo{author}{G.~S. Jeon}, \bibinfo{author}{S.-A. Kim},
  \bibinfo{author}{S.~Furukawa}, \bibinfo{author}{Y.~B. Kim},
  \bibinfo{author}{S.~Lee}, \bibinfo{author}{V.~Kiryukhin},
  \bibinfo{author}{{others}},
\newblock \bibinfo{title}{Spin {Wave} {Measurements} over the {Full}
  {Brillouin} {Zone} of {Multiferroic} {BiFeO}3},
\newblock \bibinfo{journal}{Physical Review Letters} \bibinfo{volume}{108}
  (\bibinfo{year}{2012}) \bibinfo{pages}{077202}.
\bibitem[{Skomski(2008)}]{skomski_simple_2008}
\bibinfo{author}{R.~Skomski}, \bibinfo{title}{Simple models of magnetism},
  Oxford graduate texts, \bibinfo{publisher}{Oxford University Press},
  \bibinfo{address}{Oxford ; New York}, \bibinfo{year}{2008}.
\bibitem[{Muhlbauer et~al.(2009)Muhlbauer, Binz, Jonietz, Pfleiderer, Rosch,
  Neubauer, Georgii, and Boni}]{muhlbauer_skyrmion_2009}
\bibinfo{author}{S.~Muhlbauer}, \bibinfo{author}{B.~Binz},
  \bibinfo{author}{F.~Jonietz}, \bibinfo{author}{C.~Pfleiderer},
  \bibinfo{author}{A.~Rosch}, \bibinfo{author}{A.~Neubauer},
  \bibinfo{author}{R.~Georgii}, \bibinfo{author}{P.~Boni},
\newblock \bibinfo{title}{Skyrmion {Lattice} in a {Chiral} {Magnet}},
\newblock \bibinfo{journal}{Science} \bibinfo{volume}{323}
  (\bibinfo{year}{2009}) \bibinfo{pages}{915--919}.
\bibitem[{Fert et~al.(2013)Fert, Cros, and Sampaio}]{fert_skyrmions_2013}
\bibinfo{author}{A.~Fert}, \bibinfo{author}{V.~Cros},
  \bibinfo{author}{J.~Sampaio},
\newblock \bibinfo{title}{Skyrmions on the track},
\newblock \bibinfo{journal}{Nature Nanotechnology} \bibinfo{volume}{8}
  (\bibinfo{year}{2013}) \bibinfo{pages}{152--156}.
\bibitem[{Dzyaloshinsky(1958)}]{dzyaloshinsky_thermodynamic_1958}
\bibinfo{author}{I.~Dzyaloshinsky},
\newblock \bibinfo{title}{A thermodynamic theory of “weak” ferromagnetism
  of antiferromagnetics},
\newblock \bibinfo{journal}{Journal of Physics and Chemistry of Solids}
  \bibinfo{volume}{4} (\bibinfo{year}{1958}) \bibinfo{pages}{241--255}.
\bibitem[{Moriya(1960)}]{moriya_anisotropic_1960}
\bibinfo{author}{T.~Moriya},
\newblock \bibinfo{title}{Anisotropic {Superexchange} {Interaction} and {Weak}
  {Ferromagnetism}},
\newblock \bibinfo{journal}{Physical Review} \bibinfo{volume}{120}
  (\bibinfo{year}{1960}) \bibinfo{pages}{91--98}.
\bibitem[{Rohart and Thiaville(2013)}]{rohart_skyrmion_2013}
\bibinfo{author}{S.~Rohart}, \bibinfo{author}{A.~Thiaville},
\newblock \bibinfo{title}{Skyrmion confinement in ultrathin film nanostructures
  in the presence of {Dzyaloshinskii}-{Moriya} interaction},
\newblock \bibinfo{journal}{Physical Review B} \bibinfo{volume}{88}
  (\bibinfo{year}{2013}).
\bibitem[{Katsura et~al.(2005)Katsura, Nagaosa, and
  Balatsky}]{katsura_spin_2005}
\bibinfo{author}{H.~Katsura}, \bibinfo{author}{N.~Nagaosa},
  \bibinfo{author}{A.~V. Balatsky},
\newblock \bibinfo{title}{Spin current and magnetoelectric effect in
  noncollinear magnets},
\newblock \bibinfo{journal}{Physical Review Letters} \bibinfo{volume}{95}
  (\bibinfo{year}{2005}) \bibinfo{pages}{057205}.
\bibitem[{Mostovoy(2006)}]{mostovoy_ferroelectricity_2006}
\bibinfo{author}{M.~Mostovoy},
\newblock \bibinfo{title}{Ferroelectricity in spiral magnets},
\newblock \bibinfo{journal}{Physical Review Letters} \bibinfo{volume}{96}
  (\bibinfo{year}{2006}) \bibinfo{pages}{067601}.
\bibitem[{Tuckerman et~al.(1992)Tuckerman, Berne, and
  Martyna}]{tuckerman_reversible_1992}
\bibinfo{author}{M.~Tuckerman}, \bibinfo{author}{B.~J. Berne},
  \bibinfo{author}{G.~J. Martyna},
\newblock \bibinfo{title}{Reversible multiple time scale molecular dynamics},
\newblock \bibinfo{journal}{The Journal of Chemical Physics}
  \bibinfo{volume}{97} (\bibinfo{year}{1992}) \bibinfo{pages}{1990--2001}.
\bibitem[{Daw et~al.(1993)Daw, Foiles, and Baskes}]{daw_embedded-atom_1993}
\bibinfo{author}{M.~S. Daw}, \bibinfo{author}{S.~M. Foiles},
  \bibinfo{author}{M.~I. Baskes},
\newblock \bibinfo{title}{The embedded-atom method: a review of theory and
  applications},
\newblock \bibinfo{journal}{Materials Science Reports} \bibinfo{volume}{9}
  (\bibinfo{year}{1993}) \bibinfo{pages}{251--310}.

\end{thebibliography}







\appendix

\section{Definition of the spin Hamiltonian}\label{app:hamiltonian}

In the equations of motion (eq.\ref{p_advance_T}), the mechanical
force coming from the spins can be computed from the partial
derivation of a spin Hamiltonian.  This Hamiltonian represents the
total energy of the magnetic systems.  Its general form can be
separated according to the following expression~:

\begin{equation}
  \mathcal{H} = \mathcal{H}_{z} + \mathcal{H}_{ex}+ \mathcal{H}_{an}+ \mathcal{H}_{N\acute{e}el}+ \mathcal{H}_{dm}+ \mathcal{H}_{me} +\mathcal{H}_{di},
\end{equation}

with $\mathcal{H}_{z}$ and $\mathcal{H}_{ex}$ the Zeeman and
the exchange interactions defined in Section~\ref{sec:dynamics}, the
magnetic anisotropy (2 terms), the Dzyaloshinskii-Moriya, the
magneto-electric, and the dipolar interaction.  The subsections of
this appendix give a definition of these interactions.

\subsection{Parametrization of the exchange interaction}\label{app:exchange}

Because of the celebrated Bohr-van Leuween theorem
\cite{van_leeuwen_problemes_1921}, the flow of permanent magnetic
moments cannot be the origin of the magnetism found in actual
materials and more surprisingly, magnetism is an inherently quantum
mechanical effect.  It is the interplay of electronic properties which
appear unrelated to magnetism, the Pauli principle in combination with
the Coulomb repulsion as well as the hopping of electrons that leads
to an effective coupling between the magnetic moments in a solid
\cite{duck_pauli_1998}.  This mechanism does not have a classical
analogue and is responsible for both the volume of matter and
ferromagnetism.

In SD-MD, an effective parametrization of this exchange interaction
comes with two consequences : First, it is assuming that its intensity
is rapidly decreasing with very few oscillations of its sign, so that
a rigid cutoff radius $R_{c}$ can be safely introduced.  For the
computation of the exchange interaction with a given spin $i$, only
spins $j$ such that $r_{ij} < R_{c}$ have to be taken into
account.  Second, the value of its intensity can be approximated by a
simple continuous isotropic function ${J}\left( r_{ij}
\right)$.  For $3d$ atoms, this function is based on the Bethe--Slater
curve \cite{kaneyoshi_introduction_1992, yosida_theory_1996}, and is
parametrized via three coefficients, $\alpha$ in eV, $\delta$ in
$\angstrom$, and $\gamma$ a non-dimensional constant. One has~:

\begin{equation}
  {J}\left( r_{ij} \right) = 4 \alpha \left( \frac{r_{ij}}{\delta}  \right)^2 \left( 1 - \gamma \left( \frac{r_{ij}}{\delta}  \right)^2 \right) e^{-\left( \frac{r_{ij}}{\delta}  \right)^2 }\Theta (R_c - r_{ij}), \label{BSfunction}
\end{equation}
with $\Theta (R_c - r_{ij})$ the Heaviside step function.

\begin{figure}[ht]
\centering
\includegraphics[width=0.95\columnwidth]{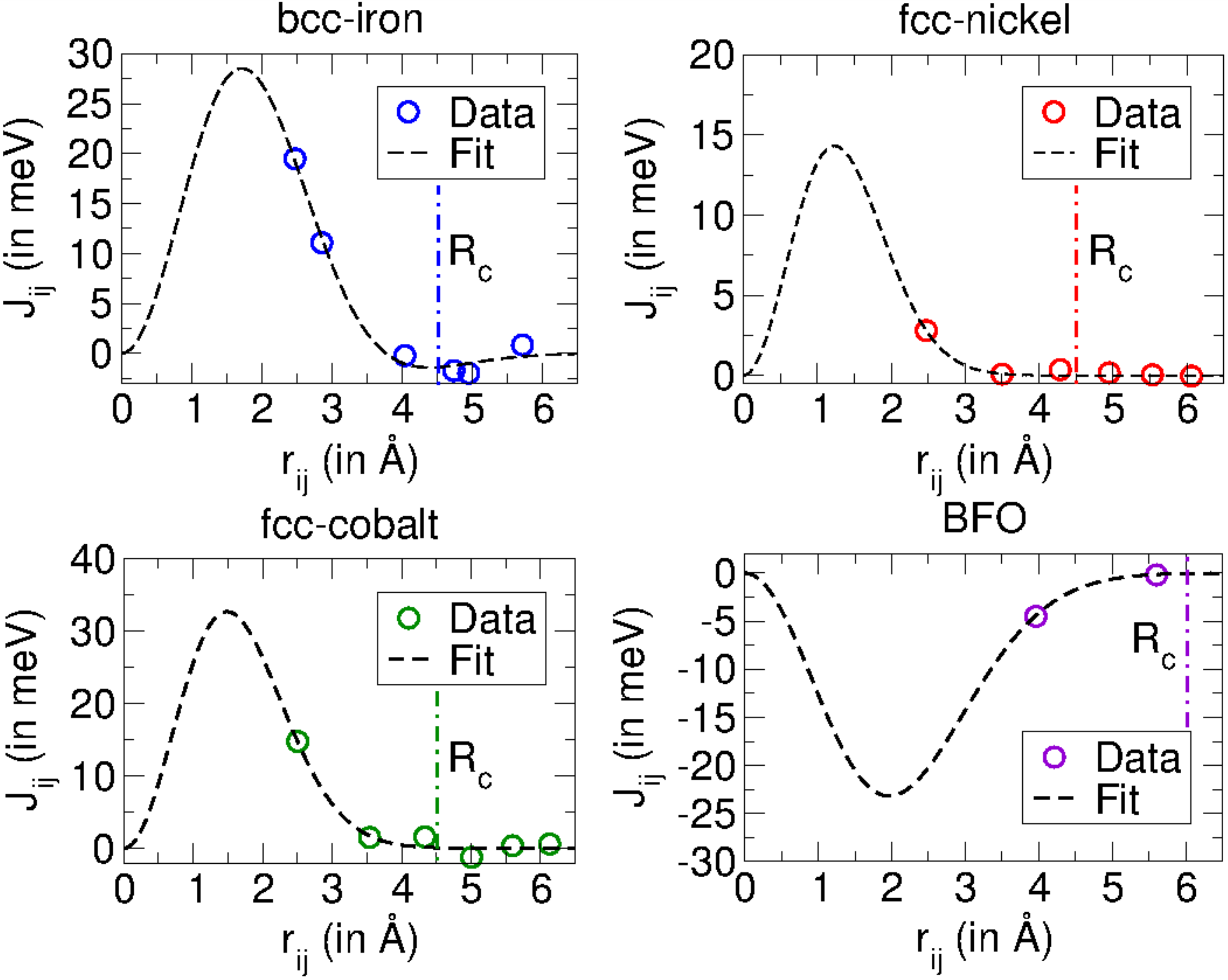}
\caption{Examples of exchange curves for bcc-Fe in blue, fcc-Ni in
  red, fcc-Co in green, and bismuth ferrite oxide (BFO) in purple. The
  dotted line labeled R$_{c}$ represents a cutoff radius taking
  into account the three first neighbor shells.} \label{fig:exchange}
\end{figure}

As examples, Fig.~\ref{fig:exchange} plots the interpolation by the
function presented in Eq.~\ref{BSfunction} of different exchange data.
On one side, the three first sets are related to pure ferromagnetic
metals such as Iron, Nickel and Cobalt, and come from \emph{ab initio}
calculations performed within the non-relativistic spin-polarized
Green-function technique \cite{pajda_ab_2001}, in general agreement
with scattering experiments.  On the other side, the fourth set is for
anti-ferromagnetic BiFeO$_3$ (BFO) and was measured by inelastic
neutron scattering technique \cite{jeong_spin_2012}.  The values of
the coefficients $\alpha$, $\gamma$ and $\delta$ to use for each of
these four materials are gathered in table~\ref{tab:exchange}.

\begin{table}[ht]
\begin{center}
\begin{tabular}{|c|c|c|c|}
  \hline
  ~          & $\alpha$ (meV) & $\gamma$& $\delta$ ($\angstrom$) \\
  \hline
  \hline
  Bcc-Fe   & 25.498            & 0.281               & 1.999 \\
  \hline
  Fcc-Ni & 9.73              & $1.08\cdot 10^{-4}$ & 1.233 \\
  \hline
  Fcc-Co & 22.213            & $8.08\cdot 10^{-6}$ & 1.485 \\
  \hline
  BFO        & -15.75            & 0                   & 1.965 \\
  \hline
\end{tabular}
\caption {Coefficients fitting the \emph{ab initio} and experimental
  results (with the model presented in Eq.~\ref{BSfunction}) for
  bcc-Fe, fcc-Ni, fcc-Co and bismuth ferrite oxide
  (BFO). \label{tab:exchange} }
\end{center}
\end{table}

\subsection{Spin orbit coupling}\label{subapp1:soc}

The spin-orbit coupling is of fundamental importance when SD-MD
simulations are at stake.  This is, for example, the case with
magnetocrystalline anisotropy, or the Dzyaloshinskii-Moriya
interaction \cite{yosida_theory_1996}.  Even if its non-relativistic
origin corresponds to energies that are orders of magnitude lower than
the exchange interaction, many technological application rely on
effective interactions arising from that spin-orbit coupling
\cite{skomski_simple_2008}.

Depending on the crystalline lattice of the material under study,
different forms of local magnetic anisotropies can arise.  A first
simple phenomenological model accounting for the one-site
magnetocrystalline anisotropy consists of shaping the angular
dependence of the corresponding energy surface with spherical
harmonics.

As an example, for uniaxial anisotropy, one has:

\begin{equation}
  \mathcal{H}_{an}= -\sum_{{ i}=1}^{N} K_{an}(\bm{r}_{
    i})\, \left( \bm{s}_{i} \cdot \bm{n}_{i} \right)^2,
\end{equation}

with $\bm{n}_{i}$ the direction of the anisotropy axis for the
spin $i$, and $K_{an}(\bm{r}_{i})$ its anisotropic constant (in
eV), which depends eventually to the position of the spin i.  With
this form of interaction, and depending on the sign of
$K_{an}(\bm{r}_{i}) $, the result can be either an easy axis or an
easy plane for the magnetization (easy axis if $K_{an}(\bm{r}_{i})
> 0$, easy plane for $K_{an}(\bm{r}_{i}) < 0$).

However, in most magnetic crystals, the magnetocrystalline anisotropy
takes more complex forms (like cubic anisotropy for example).
Besides, this anisotropy model does not exhibit a clear dependence on
the lattice parameters.

Another, more sophisticated model aimed at taking into account the
magnetocrystalline anisotropy is the two-site non-local Néel
anisotropy \cite{neel_approche_1954}.  Limited to the pseudo-dipolar
term only, one has:

\begin{eqnarray}
\mathcal{H}_{N\acute{e}el}&=&-\sum_{{ i,j=1,i\neq j}}^Ng_1(r_{ij})\left(({\bm e}_{ij}\cdot {\bm s}_{i})({\bm e}_{ij}\cdot {\bm s}_{j})-\frac{1}{3}{\bm s}_{i}\cdot{\bm s}_{j}\right)\nonumber,
\end{eqnarray}

with $g_1(r)$ a fast decreasing function of $r$. Terms like
$\sum_{{ i,j=1,i\neq j}}^Ng_1(r_{ij}){\bm s}_{i}\cdot{\bm
  s}_{j}$ can be viewed as special case of exchange interaction
and can be omitted, by considering a well-suited redefinition of
$J({r_{ij}})$.  In our implementation, $g_1(r)$ was chosen to be
the same function of three parameters as for the exchange
interaction (see eq.~\ref{BSfunction}).

It is also well known that the combination of the exchange interaction
and the spin-orbit coupling can give rise to non-collinear spin
states.  First an object of theoretical work, this canted 
ferromagnetism (the ground state presents spins that are not perfectly 
aligned, but slightly tilted with one another)
has since become a promising effect for many potential applications
\cite{muhlbauer_skyrmion_2009,fert_skyrmions_2013}.

The most common way to simulate this effect is to couple the exchange
interaction (see Section.~\ref{subsec1:dynamics} and
\ref{app:exchange}) to another interaction, referred to as the
anti--symmetric Dzyaloshinskii-Moriya interaction
\cite{dzyaloshinsky_thermodynamic_1958,moriya_anisotropic_1960}.  This
interaction takes the following form:

\begin{equation}
  \mathcal{H}_{dm} = \sum_{{ i,j}=1,i\neq j}^{N} \bm{D}(r_{ij})\cdot\left(\bm{s}_{i}\times \bm{s}_{j}\right),
\end{equation}

with $\bm{D}(r_{ij})$ the Dzyaloshinskii-Moriya vector, which
defines both the intensity (in eV) and the direction for the effect.
In particular, the Dzyaloshinskii-Moriya interaction is known to be a
key mechanism in the stabilization of magnetic skyrmions
\cite{rohart_skyrmion_2013}.

\subsection{Magneto--electric interaction}\label{subapp1:me}

In some materials, like BFO, magnetism coexists and interplays with
dielectric permanent polarization at the atomic scale.  According to
Katsura \emph{et al.} \cite{katsura_spin_2005} and Mostovoy
\cite{mostovoy_ferroelectricity_2006}, the effects of this interplay
can be taken into account within SD-MD via an anti--symmetric
spin--spin effective interaction, as a particular case of the
Dzyaloshinskii-Moriya vector, such as:

\begin{equation}
  \mathcal{H}_{me} =\sum_{i,j=1,i\neq j}^{N} \left( \bm{E}\times\bm{e}_{ij} \right) \cdot\left(\bm{s}_{i}\times \bm{s}_{j} \right),
\end{equation}

with $\bm{E}$ giving the direction and the intensity of a screened
dielectric atomic polarization (in eV).  This polarization can also be
induced by an external electric field of sufficient intensity.

\subsection{Dipolar interaction}\label{subapp1:dipolar}

With larger atomic systems, the long-range dipolar interaction is
responsible for the stabilization of domain walls and the nucleation
of magnetic domains.  As such, it is often referred to as
a demagnetizing field.

Its general non-local expression is

\begin{equation}
  \mathcal{H}_{di} = -\frac{\mu_0 \mu_B^2}{4\pi}\sum_{
    i,j=1,i\neq j}^{N} \frac{g_{i} g_{j}}{\bm{r}_{
      ij}^3} \left( (\bm{s}_{i}\cdot\bm{e}_{ij} ) (\bm{s}_{
    j}\cdot\bm{e}_{ij} ) -\frac{1}{3}(\bm{s}_{
    i}\cdot\bm{s}_{i} ) \right),
\end{equation}

with $g_{i}$ and $g_{j}$ the Land\'e factors for spins $i$ and
$j$ respectively, $\bm{r}_{ij}=\bm{r}_{j}-\bm{r}_{i}$, and
$\bm{e}_{ij}=\bm{r}_{ij}/|\bm{r}_{ij}|$.

Despite its simple formula, its computational cost is one of the main
limiting factors for large magnetic simulations.  Indeed, due to its
long-range nature, the dipolar interaction does not have a finite
cutoff distance, and therefore scales computationally as $N^2$, with
$N$ the number of atoms in the system.

To compute the dipolar effective
field and force, two solutions have been considered.  The first one
takes advantage of the properties of the Suzuki--Trotter decomposition
to avoid computing the dipolar interaction at each timestep.  In MD
simulations, this solution is often considered when thermostats or
barostats have very different characteristic time scales, and is
referred to as the r-RESPA algorithm \cite{tuckerman_reversible_1992}.
A second one relies on an assumption of periodicity in space to use
the well known technique of Ewald sums
\cite{frenkel_understanding_2002}.  For now, neither of these
solutions are implemented for the SD-MD package in LAMMPS, but are
being actively tested.  Note that because the intensity of the dipolar
interaction is usually much smaller than other considered magnetic
interactions, in magnetic systems below the paramagnetic limit that
are small enough to avoid the nucleation of domain walls, this effect
can be safely omitted.


\section{Parametrization of a mechanical potential}\label{app:mechpotential}

In MD, the standard approach to simulate properties of metals is to
consider a mechanical interaction between the magnetic atoms using a
empirial many-body potential such as the embedded atom method (EAM)
\cite{daw_embedded-atom_1993}.

Dudarev {\it et al} followed this strategy and developed a very
specific semi-empirical many-body interatomic potential suitable for
large scale molecular dynamics simulations of magnetic $\alpha$-iron
\cite{dudarev_`magneticinteratomic_2005}.  The functional form of the
embedding portion of that potential is derived using a combination of
the Stoner and the Ginzburg–Landau models, and reveals the spontaneous
magnetization of atoms by a broken symmetry of the solutions of the
Ginzburg–Landau model.  Even if this strategy provides a link between
magnetism and interatomic forces, the anisotropy effect through the
spin-orbit coupling does not emerge naturally.  This is not the
strategy we have followed in this paper because we consider the
magnitude of each atom's magnetic moment to be fixed during the
simulation. Only its direction changes over time.  However, there is
no conceptual difficulty with implementing the more detailed potential
in LAMMPS.

Because EAM potentials are either fitted to experimental or {\it ab
  initio} data, the influence of the magnetic interactions is already
silently included.  In the framework of {\it ab initio} derived EAM
potentials, this can be seen in the special case of collinear
magnetism.  Thus the standard approach is to effectively subtract the
magnetic interactions from the mechanical potential, as~:

\begin{equation}
  \mathcal{H}_{mech}^{eff}(\bm{r}_{i}) = \mathcal{H}_{mech}(\bm{r}_{i}) - \mathcal{H}_{mag}^{ground}(\bm{r}_{i}),
  \label{app:mech1}
\end{equation}

with $\mathcal{H}_{mech}^{eff}(\bm{r}_{i})$ the mechanical
potential that needs computing, $\mathcal{H}_{mech}(\bm{r}_{
  i})$ the EAM potential fitted before the magnetic energy
subtraction, and $\mathcal{H}_{mag}^{ground}(\bm{r}_{i})$
the magnetic ground-state energy value.  For example in a ferromagnet,
because the exchange energy is by far the most intense value, the
associated ground state is given by the following sum, with $j$ the
neighboring atoms of the atom $i$:
\begin{equation}
  \mathcal{H}_{mag}^{ground}(\bm{r}_{i}) = \sum_{{ j}=1}^{Neigh} {J}(r_{ij})\, |\bm{s}_{i}|\, |\bm{s}_{j}|. 
\end{equation}

Future work will study how the combination of both experimental and
\emph{ab initio} results can be used to derive better empirical
magneto-mechanical potentials.

\end{document}